\documentclass[PAPER, atlasdraft=false, texlive=2023, UKenglish,coverpage=false,cernpreprint, texmf, orcidlogo]{atlasdoc}
 
\usepackage{atlaspackage}
\usepackage{atlasbiblatex}
 
\usepackage{atlasphysics}

\graphicspath{{logos/}{figures/}}
 
\usepackage{ANA-EXOT-2019-06-PAPER-defs}

\AtlasTitle{Search for pair-production of vector-like quarks in lepton+jets final states containing at least one $b$-tagged jet using the Run 2 data from the ATLAS experiment}

\AtlasAbstract{
A search is presented for the pair-production of heavy vector-like quarks in the lepton+jets final state using 140~fb$^{-1}$ of proton--proton collisions at $\sqrt{s}= 13$~TeV collected with the ATLAS detector. The search is optimised for vector-like top-quarks ($T$) that decay into a $W$ boson and a $b$-quark, with one $W$ boson decaying leptonically and the other hadronically. Other vector-like quark flavours and decay modes are also considered. Events are selected with one high transverse-momentum electron or muon, large missing transverse momentum, a large-radius jet identified as a $W$ boson, and multiple small-radius jets, at least one of which is $b$-tagged. Vector-like $T$-quarks with 100\% branching ratio to $Wb$ are excluded at 95\% CL for masses below 1700~GeV. These limits are also applied to vector-like $Y$-quarks, which decay exclusively into a $W$ boson and a $b$-quark. Isospin singlets with $\ensuremath{ {\cal B}(T \to Wb:Ht:Zt)}={1/2}:{1/4}:{1/4}$ are excluded for masses below 1360~GeV.
}

\AtlasRefCode{EXOT-2019-06}
 
\PreprintIdNumber{CERN-EP-2023-254}

\AtlasJournalRef{Phys.\ Lett.\ B. 854 (2024) 138743 }
\AtlasDOI{10.1016/j.physletb.2024.138743}

\hypersetup{pdftitle={ATLAS document},pdfauthor={The ATLAS Collaboration}}
\usepackage{makecell}
\usepackage{tikz}
\usepackage{bm}
\usepackage{multirow}
\begin{document}
 
\maketitle

\section{Introduction}
\label{sec:intro}

\begin{figure}[h]
\centering
\subfloat[]{
\includegraphics[width=0.45\textwidth]{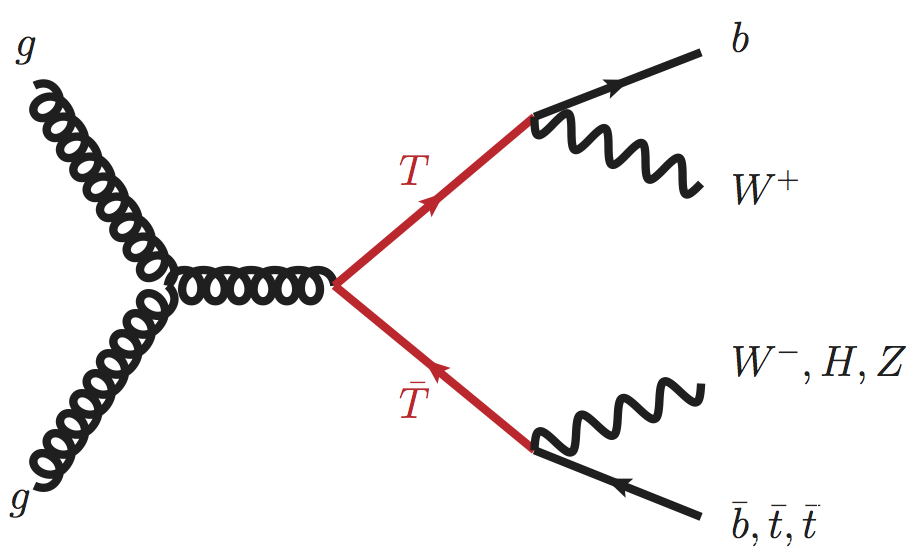}
\label{fig:Feyn_a}
}
\subfloat[]{
\includegraphics[width=0.45\textwidth]{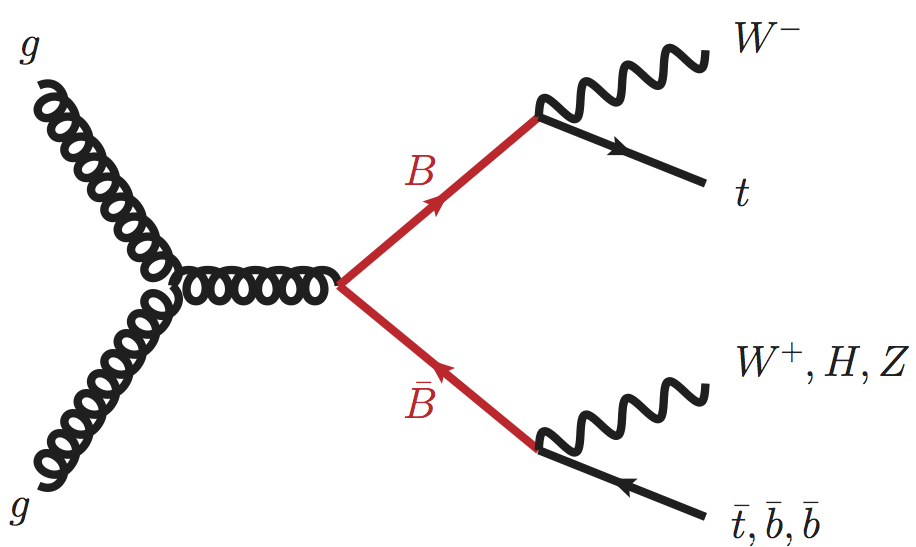}
\label{fig:Feyn_b}
}
\caption{Representative tree-level Feynman diagrams for \protect\subref{fig:Feyn_a} \TTbar{} and \protect\subref{fig:Feyn_b} \BBbar{} production for which at least one of the VLQs decays into a $W$ boson.}
\label{fig:Feyn}
\end{figure}

The Standard Model (SM) of particle physics is extremely successful in describing elementary particles and their interactions, yet
its many shortcomings reveal that it is incomplete.
In particular, naturalness~\cite{Susskind:1978ms} suggests that a more complete theory will provide an explanation for how radiative divergences to the Higgs boson mass from $t$-quark loops are cancelled out.
Extensions of the SM, such as extra dimensions~\cite{Randall:1999ee}, composite Higgs bosons~\cite{Kaplan:1983sm,Vignaroli:2012sf} and Little Higgs boson~\cite{Schmaltz:2005ky} models, predict the existence of vector-like quarks (VLQs) that could mitigate these large radiative corrections to the Higgs boson mass.
The VLQs are spin-1/2, colour-triplets that have the same weak isospin for left- and right-handed chiralities.
Thus, VLQs would not acquire mass via the Higgs boson~\cite{Aguilar-Saavedra:2013qpa}, allowing them to evade the limits that exclude additional SM-like quarks~\cite{Eberhardt:2012gv}.
To cancel out the Higgs boson mass divergence from top-quark loops, the VLQs must couple preferentially to third-generation quarks.
VLQs could appear as different types of multiplets: SU(2) singlets, doublets, or triplets of $T$, $B$, $X$ or $Y$; where $T$ and $B$ have the same electric charge as the SM $t$- and $b$-quarks, while $X$ and $Y$ have electric charges $5/3$ and  $-4/3$, respectively.
In the simplest models, the mass difference between VLQs in a given SU(2) multiplet must be small to satisfy constraints from precision electroweak measurements~\cite{Aguilar-Saavedra:2013qpa}, excluding cascade decays such as $T \to WB \to WWt$.
The VLQs can decay via a flavour-changing neutral current or a
charged current, allowing the $T$ and $B$ to each have three possible
decays: $T \to Wb/Zt/Ht$ and $B \to Wt/Zb/Hb$. The  $X$ and $Y$ have no SM partners, so these can only decay via $X \to Wt$ and  $Y \to Wb$.
Decays into final states with first and second generation quarks, though not forbidden, are not favoured as they would not address the hierarchy
problem.
Examples of Feynman diagrams for $T$ and $B$ pair production and decay are shown in Figure~\ref{fig:Feyn}.
 
The branching ratio of the VLQ decay to SM particles is not fixed by the theory, therefore, all possible VLQ decays and branching ratios need to be probed in searches for VLQs.
A combination of searches for pair-production of VLQs by ATLAS using
\SI{36}{\ifb} of Run~2 data collected at the Large Hadron Collider (LHC) from 2015 to 2016 at a center-of-mass energy of $\sqrt{s}= 13$~\TeV{} excludes VLQs with masses
below \SI{1310}{\GeV} for any branching ratio to the assumed decays~\cite{EXOT-2017-17}.
ATLAS searches for pair-produced VLQs using the full proton--proton (\pp{}) collision data sample of \SI{140}{\ifb} collected during Run~2 of the LHC between 2015 and 2018 at $\sqrt{s}= 13$~\TeV{} include a search in final states with one lepton, jets and large missing transverse momentum~\cite{atlascollaboration2022search}, mostly sensitive to the decay $T\rightarrow Zt$ or $B \rightarrow Wt$, and in final states with at least one leptonically decaying $Z$ boson and a third-generation quark~\cite{2023138019}, which is mainly sensitive to $T \rightarrow Zt$ and $B \rightarrow Zb$.
CMS published searches for pair production~\cite{CMS-B2G-19-005,CMS:2022fck} of vector-like quarks using \SI{137}{\ifb} of data collected during the LHC Run 2.
 
ATLAS has previously performed a search for pair-produced VLQs in the one lepton+jets final state using \SI{36}{\ifb} of Run~2 data which was optimised to the $T \rightarrow Wb$ decay~\cite{EXOT-2016-14}.
That search excluded $T$-quark masses below \SI{1350}{\GeV} in the scenario $\BR{T\to Wb} = 1$ and $T$-quark masses below \SI{1170}{\GeV} in the $T$ isospin singlet scenario with a branching ratio of $\BR{T \to Wb:Ht:Zt}={1/2}:{1/4}:{1/4}$.
The recent search for pair-produced VLQs by CMS in leptonic final states using the full CMS Run~2 data sample excludes the scenario $\BR{T\to Wb} = 1$ for masses below \SI{1540}{GeV}~\cite{CMS:2022fck}.
 
Singly produced VLQs have also been searched for by ATLAS~\cite{Aad:2858026,Aad_2022} and CMS~\cite{CMS-B2G-19-004,CMS-B2G-20-010,CMS:2023agg} using the full LHC Run~2 data sample, but the interpretation of the search results depends on an additional constant for the coupling to electroweak bosons~\cite{Roy_2020}.
 
This paper presents a search for the pair production of VLQs decaying
into third-generation quarks using the \pp{} collision data collected
at the LHC from 2015 to 2018 at a
centre-of-mass energy of \SI{13}{\TeV} with the ATLAS experiment, increasing the data sample relative to the previous search in this decay channel from \SI{36}{\ifb} to \SI{140}{\ifb}.
The analysis is optimised for the $\TTbar \to WbW\bar{b}$ channel with one
$W$ boson decaying leptonically and the other hadronically, resulting in a final state with exactly one lepton, missing transverse momentum and jets, but the search is also sensitive to the other VLQs and decay modes.
Targeting events with a leptonically decaying $W$ boson ($W\to \ell\nu$ with
$\ell=e,\mu$) suppresses SM processes with purely
hadronic final states, while the hadronically decaying $W$ boson provides a
large branching ratio.
Vector-like $T$ candidates are reconstructed such that the mass
difference between the leptonically and hadronically decaying $T$
candidates is minimised.
The reconstructed mass of the leptonically decaying $T$ is then used as the
discriminating variable to test for the presence of a VLQ signal.
The dominant background processes are \ttbar{}, \Wjets{}, and
single-top-quark production.
Control regions enhanced in \ttbar{} or \Wjets{} events are used to
correct the mismodelling observed in Monte-Carlo (MC) simulations of those processes.
Finally, a profile likelihood fit is performed on the observed distributions of the reconstructed VLQ mass to test for the presence of VLQ signals for various signal models, varying the VLQ mass and the decay branching ratios.
The data-driven corrections to \ttbar{} and \Wjets{} in dedicated control regions are new with respect to previous analyses.
The statistical analysis also includes additional control regions to better constrain the dominant backgrounds characterised by substantial modeling uncertainties, while maintaining a high signal acceptance and achieving a narrow peak in the reconstructed VLQ mass distribution.
Further improvements are brought by the improved object identification, especially from $W$-boson tagging.


\section{ATLAS detector}
\label{sec:detector}


\newcommand{\AtlasCoordFootnote}{
ATLAS uses a right-handed coordinate system with its origin at the nominal interaction point (IP)
in the centre of the detector and the \(z\)-axis along the beam pipe.
The \(x\)-axis points from the IP to the centre of the LHC ring,
and the \(y\)-axis points upwards.
Polar coordinates \((r,\phi)\) are used in the transverse plane,
\(\phi\) being the azimuthal angle around the \(z\)-axis.
The pseudorapidity is defined in terms of the polar angle \(\theta\) as \(\eta = -\ln \tan(\theta/2)\).
Angular distance is measured in units of \(\Delta R \equiv \sqrt{(\Delta\eta)^{2} + (\Delta\phi)^{2}}\).}

The ATLAS experiment~\cite{PERF-2007-01} at the LHC is a multipurpose particle detector
with a forward--backward symmetric cylindrical geometry and a near \(4\pi\) coverage in
solid angle.\footnote{\AtlasCoordFootnote}
It consists of an inner tracking detector (ID) surrounded by a thin superconducting solenoid
providing a \SI{2}{\tesla} axial magnetic field, electromagnetic and hadron calorimeters, and a muon spectrometer.
The inner tracking detector covers the pseudorapidity range \(|\eta| < 2.5\).
It consists of silicon pixel, silicon microstrip, and transition radiation tracking detectors.
Lead/liquid-argon (LAr) sampling calorimeters provide electromagnetic (EM) energy measurements with high granularity.
A steel/scintillator-tile hadron calorimeter covers the central pseudorapidity range (\(|\eta| < 1.7\)).
The endcap and forward regions are instrumented with LAr calorimeters for both the EM and hadronic energy measurements up to \(|\eta| = 4.9\).
The muon spectrometer (MS) surrounds the calorimeters and is based on three large superconducting air-core toroidal magnets with eight coils each.
The field integral of the toroids ranges between \num{2.0} and \SI{6.0}{\tesla\metre}
across most of the detector.
The muon spectrometer includes a system of precision tracking chambers and fast detectors for triggering.
A two-level trigger system is used to select events.
The first-level trigger is implemented in hardware and uses a subset of the detector information to accept events at a rate below \SI{100}{\kHz}.
This is followed by a software-based trigger that reduces the accepted event rate to \SI{1}{\kHz} on average depending on the data-taking conditions.
An extensive software suite~\cite{ATL-SOFT-PUB-2021-001} is used in data simulation, in the reconstruction and analysis of real and simulated data, in detector operations, and in the trigger and data acquisition systems of the experiment.


\section{Data and simulated event samples}
\label{sec:samples}

The data sample analysed is from \pp{} collisions with a centre-of-mass energy of $\sqrt{s}=\SI{13}{\TeV}$.
The data were collected between 2015 and 2018 with the ATLAS detector and correspond to an integrated luminosity of
\SI{140}{\ifb}~\cite{arXiv:2212.09379,Avoni:2633501}.
A set of single-electron~\cite{TRIG-2018-05} and single-muon triggers~\cite{TRIG-2018-01} were used, with transverse momentum (\pt{}) lowest thresholds in the range of 20--\SI{26}{\GeV} depending on the lepton flavour and data-taking period.
In addition, data were recorded using a trigger targeting events with large missing transverse energy (\MET{})~\cite{TRIG-2019-01} with thresholds of 70, 90 or \SI{110}{\GeV}, depending on the data taking period.
All detector subsystems are required to be operational during data taking and to satisfy data quality requirements~\cite{DAPR-2018-01}.

Signal events and SM background with at least one prompt lepton are modelled by MC simulation.
Contributions from processes surviving selection requirements due to non-prompt leptons or hadronic jets misidentified as leptons (dominated by QCD multijet events) are estimated by a data-driven method, with MC samples serving as cross checks.
All samples were produced using the ATLAS simulation infrastructure~\cite{SOFT-2010-01} and \GEANT~\cite{Agostinelli:2002hh}.
The estimate of some systematic uncertainties used a faster detector simulation employing a parameterisation of the calorimeter response~\cite{ATL-PHYS-PUB-2010-013}.
Unless specified otherwise, the parton shower, hadronisation, and underlying events are modelled using \PYTHIA{8.230}~\cite{Sjostrand:2014zea}, with parameters set according to the A14 set of tunable parameters (the A14 ``tune'')~\cite{ATL-PHYS-PUB-2014-021} and using the \NNPDF[2.3lo] set of parton distribution functions (PDFs)~\cite{Ball:2012cx}.
In the simulation, the $t$-quark and SM Higgs boson masses were set to 172.5~\GeV and 125~\GeV, respectively.
The effect of additional \pp{} interactions per bunch crossing (pile-up) is accounted for by overlaying the hard-scattering process with minimum-bias events simulated with \Pythia[8.186]~\cite{Sjostrand:2007gs} using the A3 tune~\cite{ATL-PHYS-PUB-2016-017} and \NNPDF[2.3lo] PDFs.
The distribution of the mean number of interactions per bunch crossing in simulation is reweighted such that its distribution matches that measured in the data.

Signal events for the production and decay of \TTbar{} and \BBbar{} were simulated at leading-order using \Protos{} v.2.2~\cite{Aguilar-Saavedra:2009xmz} with VLQ masses from \SI{1000}{\GeV} to \SI{1800}{\GeV} in steps of \SI{100}{\GeV}, and with a mass of \SI{2000}{\GeV}.
The samples were produced for the singlet model, with alternative branching ratio scenarios obtained by reweighting the events based on the decay mode.
Samples for the doublet model with a mass of \SI{1.2}{\TeV} were also produced to confirm that kinematic differences due to isospin have negligible impact on the results.
The signal cross-sections were calculated with \TOPpp[2.0]~\cite{Czakon:2011xx} at next-to-next-to-leading-order (NNLO) in QCD including the resummation of next-to-next-to-leading-logarithmic (NNLL) soft-gluon terms.

The dominant background arises from the production of $t$-quark pairs (\ttbar{}).
Events with a single $t$-quark (henceforth referred to as ``single top'') or a $W$ boson produced in association with jets (\Wjets{}) also make significant contributions.
Finally, events containing a $Z$ boson and jets (\Zjets{}), three or four $t$-quarks, multi-boson events, and $t$-quark pairs produced in association with a vector boson or a Higgs boson ($ttV$ with $V=W,Z$ and $ttH$, respectively) have small contributions.
 
The production of \ttbar{} events is modelled using the \POWHEGBOX[v2]~\cite{Frixione:2007nw,Nason:2004rx,Frixione:2007vw,Alioli:2010xd} generator at NLO accuracy in QCD with the \NNPDF[3.0nlo]~\cite{Ball:2014uwa} PDFs and the \hdamp{} parameter\footnote{The \hdamp{} parameter is a resummation damping factor and one of the parameters that controls the matching of \POWHEG{} matrix elements (ME) to the parton shower and thus effectively regulates the high-\pT{} radiation against which the \ttbar{} system recoils.} set to $1.5 m_t$~\cite{ATL-PHYS-PUB-2016-020}, where $m_t$ denotes the $t$-quark mass.
The dependence on the parton shower and hadronisation models is evaluated by comparing the nominal \ttbar{} sample with an alternative sample also produced with \POWHEGBOX[v2] generator using the \NNPDF[3.0nlo] PDFs, but instead interfaced to \HERWIG{7.04}~\cite{Bahr:2008pv,Bellm:2015jjp} using the \MMHT[lo] PDFs~\cite{Harland-Lang:2014zoa} and the H7UE tune~\cite{Bellm:2015jjp}.
An uncertainty in the matching of the NLO matrix elements (ME) to the parton shower is assessed by comparing the nominal sample to one simulated with \MGNLO[2.6.0]~\cite{Alwall:2014hca} using the \NNPDF[3.0nlo] PDFs.
The effect from possibly under-estimating the initial-state radiation (ISR) is evaluated by comparing the nominal sample to one produced with \hdamp{} increased to 3$m_t$, the renormalisation and factorisation scales divided by two, and using the ``Var3cUp'' weight of the A14 tune.
The effect from possibly over-estimating the ISR is evaluated by comparing the nominal sample to one obtained by doubling the renormalisation and factorisation scales and choosing the ``Var3cDown'' weight of the A14 tune~\cite{ATL-PHYS-PUB-2020-023}. 
The size of the uncertainty in the modelling of final state radiation (FSR) is evaluated by doubling or halving the renormalisation scale for emissions from the parton shower.

The associated production of a $t$-quark with a $W$ boson ($tW$) accounts for the vast majority of single top events. These events were modelled with the \POWHEGBOX[v2] generator at NLO in QCD using the five-flavour scheme and the \NNPDF[3.0nlo] PDFs.
The nominal sample uses the diagram removal (DR) scheme~\cite{Frixione:2008yi} to remove the contribution from Feynman diagrams already included in the \ttbar{} production.
An alternative sample simulated using the diagram subtraction scheme (DS)~\cite{Frixione:2008yi,ATL-PHYS-PUB-2016-020} is used to evaluate an uncertainty due to the choice of \ttbar/$tW$ overlap removal scheme.
The uncertainty due to the parton shower and hadronisation models is evaluated by comparing the nominal sample of events with events simulated by the \POWHEGBOX[v2] generator interfaced to \HERWIG{7.04} using the H7UE tune and the \MMHT[lo] PDFs.
To assess the uncertainty in the matching of the ME to the parton shower, the nominal $tW$ sample is compared with
a sample simulated with the \MGNLO[2.6.2] generator at NLO in QCD using the five-flavour scheme and the \NNPDF[2.3lo] PDFs~\cite{Ball:2014uwa}.

The small contributions from single top $t$-channel and $s$-channel production were modelled using the \POWHEGBOX[v2]~\cite{Frederix:2012dh,Nason:2004rx,Frixione:2007vw,Alioli:2010xd,Alioli:2009je} generator at NLO in QCD using the four-flavour scheme ($t$-channel) or the five-flavour scheme ($s$-channel) and the \NNPDF[3.0nlo] PDFs.
The uncertainty due to the parton shower and hadronisation models is evaluated by comparing the nominal sample of events with events simulated by the \POWHEGBOX[v2] generator at NLO accuracy in QCD using the five-flavour scheme and interfaced to \HERWIG{7.04} using the H7UE tune and the \MMHT[lo] PDFs.
To assess the uncertainty in the matching of NLO accuracy ME to the parton shower, the nominal sample is compared
with a sample simulated with the \MGNLO[2.6.2] generator at NLO accuracy in QCD using the five-flavour scheme and \NNPDF[2.3nlo] or \NNPDF[3.0nlo] PDFs for the $t$- and $s$-channel, respectively.

The production of $V+$jets ($V=W,Z$) was simulated with the \SHERPA{2.2.1}~\cite{Bothmann:2019yzt} generator at NLO accuracy in QCD for up to two partons, and at LO accuracy for up to four additional partons.
The full samples are then normalised to the NNLO predictions~\cite{Anastasiou:2003ds}.
Samples of diboson events ($VV$) were simulated with the \SHERPA{2.2.1} or 2.2.2~\cite{Bothmann:2019yzt} generator at NLO accuracy in QCD for up to one additional parton and at LO accuracy for up to three additional parton emissions, including off-shell effects and Higgs-boson contributions, where appropriate.
The ME calculations for \SHERPA{} samples were matched and merged with the parton shower based on Catani--Seymour dipole factorisation~\cite{Gleisberg:2008fv,Schumann:2007mg} using the \MEPSatNLO{} prescription.
The virtual QCD corrections were provided by the \OPENLOOPS library~\cite{Buccioni:2019sur,Cascioli:2011va,Denner:2016kdg}.
The \NNPDF[3.0nnlo] set of PDFs was used, along with the dedicated tune developed by the \SHERPA{} authors.
 
The production of $\ttbar V$ and $tWZ$ events was modelled using the \MGNLO[2.3.3] generator at NLO with the \NNPDF[3.0nlo] PDF and \PYTHIA{8.210} (\PYTHIA{8.211}) for $\ttbar V$ ($tWZ$) with the A14 tune and the \NNPDF[2.3lo] PDFs for the parton shower and hadronisation.
The diagram removal scheme described in Ref.~\cite{Frixione:2008yi} was employed to treat the overlap between $tWZ$ and $\ttbar Z$, and was applied to the $tWZ$ sample.
The production of $\ttbar H$ events is modelled using the \POWHEGBOX[v2] generator at NLO with the \NNPDF[3.0nlo] PDF set.
The production of $tZ$, three and four $t$-quarks events, and $t\bar{t}WW$ events was simulated with \MGNLO[2.2.2] interfaced to \PYTHIA{8.186} using the A14 tune.
The small contributions from processes from $\ttbar H$, $tWZ$, $tZ$, $ttWW$ and multi-top production is summarised as ``rare top'' processes.
Multijet MC samples were simulated using the \SHERPA[2.1.1] generator.
The matrix element calculation was included for the \(2\rightarrow2\) process at leading-order, and the default \SHERPA parton shower based on Catani--Seymour dipole factorisation was used for the showering with \pT{} ordering, using the \CT[10] PDF set~\cite{Gao:2013xoa}. 
 
Decays of $b$- and $c$-hadrons were simulated by \EVTGEN[1.6.0]~\cite{Lange:2001uf} in all simulations except \Sherpa{}, for which the default \Sherpa{} configuration recommended by the \Sherpa{} authors was used.


\section{Object reconstruction}
\label{sec:object}

Events are required to contain at least one vertex with two or more associated tracks that must have $p_{T}>500$~MeV.
Among all vertices, the vertex with the highest $p_{T}^{2}$ sum of the associated tracks is taken as the primary vertex (PV)~\cite{ATL-PHYS-PUB-2015-026}.
 
Electrons are reconstructed from clusters in the EM calorimeter matched with ID tracks and are calibrated~\cite{EGAM-2018-01}.
Electron candidates must be in the central region of the detector $\left( \left| \eta \right| < 2.47 \right)$ with $\pt{} > 27$~\GeV{} and match a track with $\left| z_0 \sin \theta \right| < 0.5$~mm and $\left| d_0 / \sigma_{d_0} \right| < 5$; where $d_0$ is the impact parameter between the track and the beam line, $\sigma_{d_0}$ is the measured uncertainty in $d_0$, and $z_0$ is the minimum distance in $z$ between the track and the beam spot.
Any candidates in the transition region between the barrel and endcap calorimeters $\left( 1.37 < \left| \eta \right| < 1.52 \right)$ are removed.
``Baseline electrons'' must satisfy the \textit{medium likelihood} identification criteria~\cite{EGAM-2018-01}, with no selection on the isolation.
``Tight electrons'' must satisfy the \textit{tight likelihood} identification criteria, plus the following isolation requirements in both the calorimeter and the ID~\cite{EGAM-2018-01}.
The first isolation requirement is  $E_{\text{T, cone}}^{\text{isol}} / \pt{e} < 0.2$; where $\pt{e}$ is the electron candidate \pt{} and $E_{\text{T,cone}}^{\text{isol}}$ is the energy deposited in the calorimeter within a cone of size $R = 0.2$ around the candidate direction; any leakage energy and energy from pile-up are subtracted.
The second isolation requirement is $p_{\text{T,var}}^{\text{isol}} / \pt{e} < 0.15$; where $p_{\text{T,var}}^{\text{isol}}$ is the sum of the track \pt{} without the electron candidate in a cone of radius $R=\min(\SI{10}{GeV} / \pt{e} , 0.2)$.
Scale factors are used to correct for differences between data and simulation for reconstruction, identification, isolation and trigger selection efficiencies~\cite{EGAM-2018-01}.

Muons are reconstructed~\cite{MUON-2018-03} from combined tracks in the MS and the ID, with ``baseline muons'' required to satisfy the \textit{loose} identification criteria and no selection on the isolation, while ``tight muons'' must satisfy the \textit{tight} identification criteria~\cite{MUON-2018-03} and satisfy the track-based isolation requirements defined by the \textit{TightTrackOnly} working point.
This working point uses the scalar sum of the \pt{} of all tracks that are within a cone of size $R = \min(0.3,\SI{10}{\GeV}/\pt{\mu})$ around the muon candidate, where \pt{\mu} is the candidate muon \pt{}.
The track matched to the muon candidate under consideration is excluded from the sum.
The muon is selected if this sum is less than 6\% of \pt{\mu}.
Finally, all muon candidates are required to satisfy $|z_0 \sin\theta| < \SI{0.5}{mm}$ and $\left| d_0 / \sigma_{d_0} \right| < 3$.
Muons are calibrated~\cite{MUON-2022-01} and are required to have $\pt{}>\SI{15}{\GeV}$ and to be reconstructed within $|\eta|=2.5$.
Scale factors are used to correct for differences in muon reconstruction, identification, vertex matching, isolation and trigger efficiencies between simulation and data~\cite{MUON-2018-03}.
 
Small-radius (small-$R$) jet candidates are built from particle-flow objects~\cite{PERF-2015-09}, using the \antikt algorithm~\cite{Cacciari:2008gp,Fastjet} with a radius parameter of $R=0.4$.
The particle-flow algorithm combines information about tracks in the ID and energy deposits in the calorimeters to form the input for the jet reconstruction.
The jet energy is calibrated to the particle level scale by using a sequence of corrections, including simulation-based corrections and in situ calibrations~\cite{JETM-2018-05}.
Jets are required to have $\pt{} > \SI{25}{\GeV}$ and $|\eta| < 2.5$.
To reject jets originating from \pileup interactions, jet candidates with $|\eta| < 2.4$ and $\pt{} < \SI{60}{\GeV}$ are required to satisfy the ``tight'' jet vertex tagger (JVT) criterion~\cite{PERF-2014-03}.
An algorithm based on deep and recurrent neural networks, called DL1r, is used to identify small-$R$ jets containing a $b$-hadron decay~\cite{ATLAS:2022qxm}. Jets are considered to be \btagged{} if they satisy the criteria for the operating point with an efficiency of $77\,\%$ and mistag rates for charm and light jets of 17.7\% and 0.52\%, respectively, as determined in simulated \ttbar events.
The $b$-quark tagging efficiencies in simulation, and the charm and light jet mistag rates, are corrected to match the efficiencies in data~\cite{FTAG-2018-01,arXiv:2301.06319,FTAG-2020-08}.
 
An overlap removal procedure is applied to prevent double counting of
ambiguous reconstructed objects, using the baseline lepton
definitions.
First, electron--muon overlap is handled by removing muons sharing a track in the ID with an electron if the muon is calorimeter-tagged~\cite{MUON-2018-03}, and otherwise removing the electron.
Next, overlap between jets and leptons is removed by rejecting any jets within $\Delta R = 0.2$ of an electron, followed by the rejection of any electrons within $\Delta R = 0.4$ of the remaining jets.
Similarly, jets are discarded if they have fewer than three matched tracks and are within $\Delta R = 0.2$ of a muon candidate.
Otherwise, the muon is rejected if it lies within $\Delta R = \text{min}(0.4, 0.04 +
\SI{10}{\GeV}/\pt{\mu})$ of a jet.
 
The missing transverse momentum, with magnitude \met, is defined as the negative vectorial sum of the transverse momenta of all calibrated
objects in an event, plus a track-based soft-term which takes into
account energy depositions matched to the primary vertex but not
matched with any calibrated object~\cite{PERF-2016-07}.
 
Finally, large-radius (large-$R$) jets are constructed from the
noise-suppressed topological calorimeter-cell clusters calibrated by using local hadronic cell reweighting~\cite{PERF-2014-07} and
the \antikt{} algorithm with $R=1.0$.
To reduce the impact of soft radiation, a grooming algorithm called ``trimming''~\cite{Krohn_2010} is applied.
Constituent small-$R$ jets, reclustered with the $k_{t}$ algorithm with a radius parameter $R=0.2$, with \pt{} less than \SI{5}{\percent} of the large-$R$ jet \pt{} are removed.
The large-$R$ jets are required to have $\pt{} > \SI{200}{\GeV}$, $|\eta| < 2.0$ and a mass larger than $\SI{50}{\GeV}$.
The jet energy scale (JES) and resolution (JER) and the mass scale (JMS) and resolution (JMR) of large-radius jets are calibrated~\cite{JETM-2018-02,ATLAS-CONF-2020-022}.
A $W$ boson tagging algorithm identifies high-\pt{} hadronically decaying $W$ bosons, whose decay products are collimated to a single large-radius jet~\cite{JETM-2018-03}, with an efficiency of 80\%.
The $W$-tagging algorithm applies criteria on the mass of the large-radius jet, the number of inner-detector tracks associated with the jets and the energy correlation function ratio $D_{2}$~\cite{JETM-2018-03,Larkoski_2014}.
Scale factors correct the $W$ boson tagging efficiency in simulation to match the efficiency measured in data~\cite{ATL-PHYS-PUB-2020-017}.


\section{Event selection, reconstruction and categorisation}
\label{sec:event}

This search targets final states with exactly one electron or muon, missing transverse momentum, and jets.
Events are selected by a single lepton or \MET{} trigger, as described in Section~\ref{sec:samples}. 
If the event is selected by a single lepton trigger, the selected lepton must have $\pt{} > \SI{27}{\GeV}$ and be matched to the object that triggered the event.  If the event is selected by the \MET{} trigger, it must have $\MET{} > \SI{200}{\GeV}$.
The \MET{} trigger selection is added  to compensate for the efficiency loss of the single muon triggers at high muon \pt{}.
These lepton and the \MET{} requirements ensure full trigger efficiency over all data taking periods.

Events are required to have exactly one tight muon with $\pt{} > \SI{27}{\GeV}$ or one tight electron with $\pt{} > \SI{60}{\GeV}$.
Increasing the $\pt{}$ threshold for electrons to $\pt{} > \SI{60}{\GeV}$ facilitates the estimation of the background from non-prompt electrons.
The fake lepton background estimation relies on the fact that, compared to real leptons, fake leptons that satisfy the baseline lepton selection have a much lower efficiency for satisfying the tight lepton criteria.
The single electron trigger selection, which is part of both the baseline and tight electron selection already imposes very strict selections on the electrons with $\pt{} < \SI{60}{\GeV}$, making it difficult to define a selection tighter than the baseline electron criteria.
Studies showed that increasing the electron $\pt{}$ threshold to $\SI{60}{\GeV}$ does not impact the analysis sensitivity.
Events with additional leptons with $\pt{} > \SI{25}{\GeV}$ satisfying the baseline criteria are vetoed.
In addition, all events must have $\MET{}>\SI{60}{\GeV}$ and contain at least three small-$R$ jets, at least one of which must be $b$-tagged.
The criteria described above are referred to as the preselection.

The focus is on $T$-quarks with a mass above \SI{1.35}{\TeV}, which corresponds to the lower mass limit set by the previous analysis~\cite{EXOT-2016-14}.
The large mass of the $T$ will lead to decay products with large \pt{}, which in turn will have collimated decay products.
In particular, the hadronically decaying $W$ boson will produce a large-$R$ jet, so events are required to contain at least one $W$-tagged  large-$R$ jet.
If more than one large-$R$ jet is $W$-tagged, the one with a mass closest to the $W$-boson mass ($m_{W}=\SI{80.38}{\GeV}$) is selected as the hadronically decaying $W$ boson (\Whad) for the reconstruction of the hadronically decaying $T$-quark candidate.
The leptonically decaying $W$ boson (\Wlep{}) is reconstructed from the system of the selected lepton and reconstructed neutrino.
The $\vec{p}_T$ of the reconstructed neutrino is determined from the missing transverse momentum, and the $p_z$ of the reconstructed neutrino is calculated by assuming that the lepton-neutrino system has the invariant mass of the $W$ boson.
This leads to a quadratic equation.
When two real solutions are obtained, the one with the smaller absolute value for the reconstructed neutrino $p_z$ is used.
When the solutions are complex, a real solution is obtained by adjusting the $x$ and $y$ components of the neutrino momentum to minimise a $\chi^2$ parameter that includes the uncertainties in the neutrino and lepton momenta and $W$ boson mass.
Finally, the angular distance between the lepton and the reconstructed neutrino is required to be $\Delta R(\ell,\nu)< 0.7$.

Reconstruction of the two $T$ candidates is done by pairing each of the $W$ candidates, \Wlep{} and \Whad{}, with a small-$R$ jet.
The small-$R$ jets are selected for pairing with the candidate $W$ bosons by minimising the difference between the hadronic and leptonic reconstructed $T$ masses, $\dm \equiv \dmlong$.
If the event contains one $b$-tagged jet, only combinations that include the $b$-tagged jet are considered.
If the event contains two or more $b$-tagged jets, only combinations with the two $b$-tagged jets leading in \pt{} are considered.
Signal events are expected to have VLQs with low \pT{} and high mass, causing the jet paired with the \Whad{}, referred to as \bhad{}, to be well separated from the \Whad{}.  By contrast, \ttbar{} events will have high-\pt{} $t$-quarks that lead to a small opening angle between the resulting $W$ and $b$.
Therefore, a requirement of $\Delta R(\Whad,\bhad)>1.0$ is imposed to reduce the \ttbar{} background.
Well reconstructed signal events  should also have small values of \dm{}, with both the reconstructed $T$-quark masses \mTlep{} and \mThad{} close to the actual mass of the $T$-quark.
A similar statement can be made for \ttbar{} events, but the reconstructed variables \mTlep{} and \mThad{} are close to the $t$-quark mass $m_t$.
Production of \TTbar{} events with other decays, such as $T\to Ht/Zt$, will tend to have \mTlep{} and \mThad{} near the $T$-quark mass, but with large values of \dm{} as the VLQ reconstruction is less optimised for $t$-quarks from the $T$-quark decay.
Detector effects, final state radiation, and mis-reconstruction can also broaden the reconstructed mass peaks for signal and \ttbar{} events. 
In contrast, a smooth distribution in \dm{} is expected for background processes, such as \Wjets{} and single top production.
The variable $\dm$ is therefore used to separate regions enriched in single top background with similar kinematic properties as the single top events in the signal region.

\begin{figure}
\centering
\includegraphics[width=0.35\textwidth]{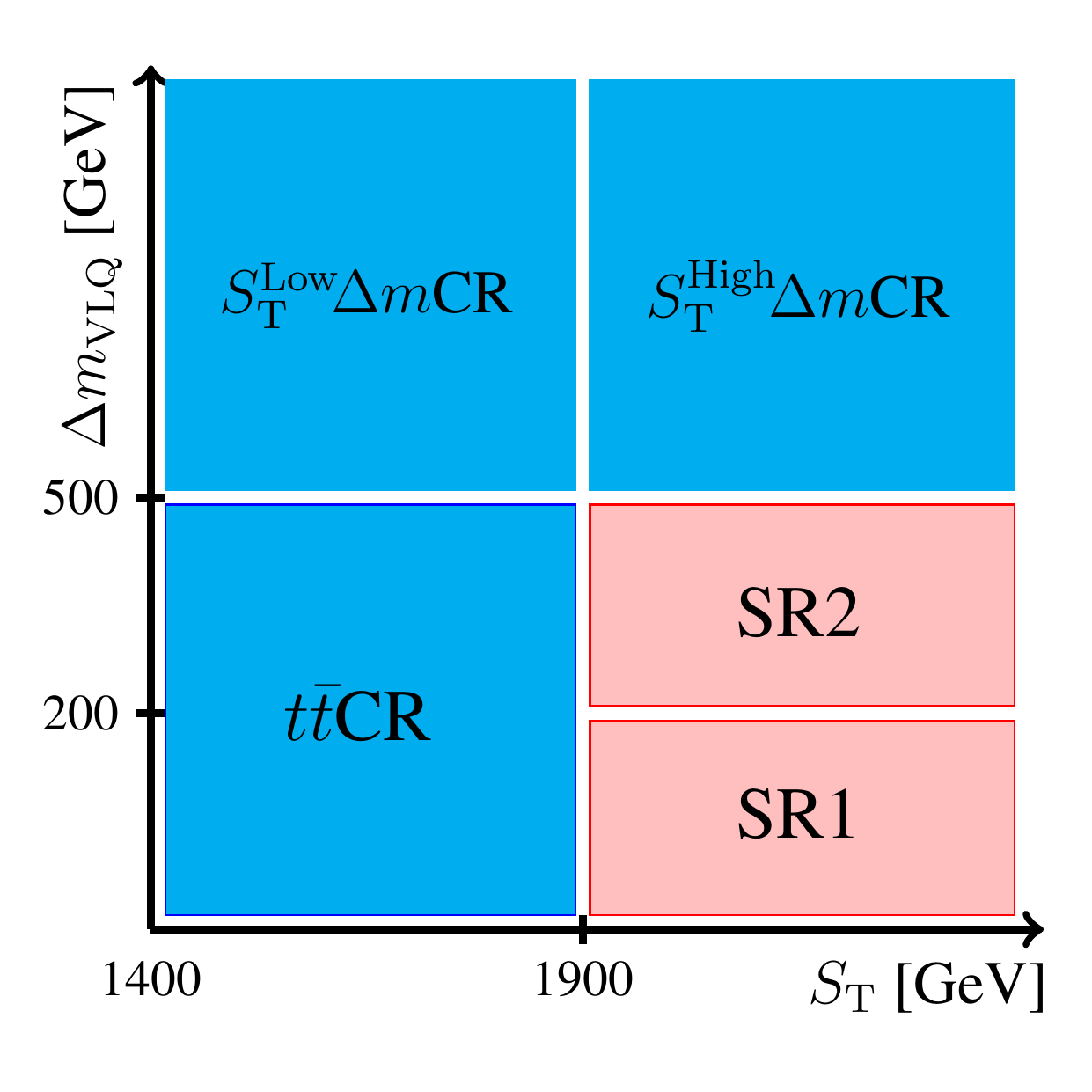}
\caption{Illustration of the two-dimensional plane in $\ST{}$ and $\dm$ on which the signal and control regions, included in the final combined likelihood fit, are defined. }
\label{fig:2Dgrid}
\end{figure}

The scalar sum $\ST{}$ of the $\pt{}$ of the selected small-$R$ jets, the lepton \pt{} and the \met, is a powerful discrimination quantity, due to the large expected mass of the $T$.
The signal regions require $\ST{}>\SI{1900}{\GeV}$, while regions to constrain the background nuisance parameters and normalisations require $1400 < \ST{} < \SI{1900}{\GeV}$.
The control and signal regions are further divided by $\dm$, as illustrated in Figure~\ref{fig:2Dgrid}.
Events in these regions are used in the final likelihood fit as described in Section~\ref{sec:result}.
The signal region is defined by $\ST{}>\SI{1900}{\GeV}$ and $\dm < \SI{500}{\GeV}$, which is further divided into the regions SR1 with $\dm < \SI{200}{\GeV}$ and SR2 with $200 < \dm < \SI{500}{\GeV}$.  The region \SRa{} is designed to capture the best-reconstructed VLQ events that would have a narrow peak in the \mTlep{} and \mThad{} distributions. The region \SRb{} will capture less-well-reconstructed VLQ events, resulting in broader mass distributions. Despite this, the fraction of VLQs expected to enter this region is non-negligible.
A control region with $\ST{} > 1900$~GeV and $\dm>\SI{500}{\GeV}$, denoted \highSTCR{}, has low signal contamination for signal masses that are not excluded yet and is dominated by single top production.
This control region provides an opportunity to constrain the modelling of the single top background using events with kinematic properties similar to the signal region.
The region with $1400 < \ST{} < 1900$~GeV is divided into two control regions in $\dm$: \ttCR{}  with $\dm<\SI{500}{\GeV}$ and \lowSTCR{} with $\dm>\SI{500}{\GeV}$.
The \ttCR{} is rich in \ttbar{} events with a purity of nearly 70\%, allowing the constraint of the normalisation and modelling of $t$-quark pair production.
The \lowSTCR{} provides a second region for constraining the single top background, but with kinematics properties similar to the \ttCR{}.
A summary of all regions is provided in Table~\ref{tab:Regions}.
The regions not included in the fit, \WCR{} and \ttRW{} (\ttbar{} reweighting region), serve to derive data-driven corrections to the background modelling and are discussed in the following section.
 
\begin{table}[htb]
\centering
\caption{Overview of the signal and control regions used in the analysis.}
\label{tab:Regions}
\scriptsize
\begin{tabular}{c | c | c | c |  c | c }
\toprule
Selection                 & SR1 / SR2       & \ttCR{}   & {\lowSTCR{} / \highSTCR} 	& \WCR{}          & \ttRW{}       \\
 
\midrule
Preselection              & $\checkmark$  & $\checkmark$    & $\checkmark$       	& $\checkmark$         & $\checkmark$    \\
$N_{\text{Large-}R\text{ Jet}}$ & $\geq 1$      &$\geq 1$         &$\geq 1$ 	& $\geq 1$             & $\geq 1$        \\
$\ST{} [\GeV]$              & $>$1900       & 1400--1900     & 1400--1900 / $>$1900 	& 900--1900             & $>$800         \\
$N_{\Wtag}$                & $\geq 1$      & $\geq 1$        & $\geq 1$  	& $\geq 1$ partially inverted    & $\geq$ 1        \\
$N_{\ensuremath{b\!-\!\text{tag}}}$     & $\geq 1$      & $\geq 1$        & $\geq 1$   	& $\geq 1$             & $\geq 2$        \\
$\Delta R(\Whad,\bhad)$  & $>1.0$        &$>1.0$           &$>1.0$	&   --                 & $<1.0$          \\
$\Delta R(\ell,\nu)$      & $<0.7$        &$<0.7$           &$<0.7$	& $<1.0$               & $<1.2$          \\
$\dm[\GeV]$                & $<200$ / 200--500	   & $<500$          & $>500$ 	& --                   & --               \\
$\mTlep, \mThad[\GeV]$             & --             & --               & -- 	& --                     & $<700$      \\
\midrule
Included in fit           & yes / yes           & yes             & yes / yes                   	& no                   & no              \\                \midrule
Goal   & \makecell{Optimise signal\\ sensitivity} & \makecell{Constrain $t\bar{t}$ \\ normalisation} & \makecell{Constrain single top \\ uncertainties} &    \makecell{Derive $W$+jets \\normalisation factor} & \makecell{Derive $t\bar{t}$ $\ST{}$ \\shape reweighting} \\
\bottomrule
\end{tabular}
 
\end{table}


\section{Background estimate}
\label{sec:bkg}

Contributions from the dominant background processes, \ttbar{} and \Wjets{} production, are estimated by using MC simulation with data-driven corrections derived in dedicated control regions dominated by the respective process.
The SM backgrounds containing a prompt-lepton (single top, \Zjets{}, diboson, etc.) are also estimated by using MC simulations.
Finally, the small contribution from multijet events is estimated by using a data-driven approach.

A correction for \Wjets{} events is derived in a dedicated control region referred to as the \WCR{}.
This region is constructed to be orthogonal to the signal region, in particular with an inverted jet mass requirement in
the hadronic $W$ boson tagging algorithm. 
Possible biases in the modelling of \Wjets{} events due to the inversion of this requirement are investigated and taken into account with a dedicated uncertainty, as detailed below.
The \WCR{} also requires $900 < \ST < \SI{1900}{GeV}$ to reduce possible contamination from VLQ signal events.
All other selection requirements are the same as in the signal region, as shown in Table~\ref{tab:Regions}.
 
Requiring at least one $b$-tagged jet, similar to the signal region requirement, causes the \WCR{} to have a significant contribution from \ttbar{} events.
This requirement helps to validate the correction to simulated $W$+jets events as the modelling of \Wjets{} depends on the heavy-flavour requirement~\cite{ATL-PHYS-PUB-2017-006}.
The estimate of the \Wjets{} correction is based on the charge asymmetry of \Wjets{} events in the \WCR{}, present due to the asymmetry in $u$- and $d$-quark content in the proton~\cite{Martin:2009iq}.
Charge asymmetry is defined as $\displaystyle A = N_+ - N_-$, where $N_+$ and $N_-$ are the numbers of events with a positively or negatively charged lepton, respectively.
The correction factor to the \Wjets{} normalisation is calculated by comparing $A$ in data and MC simulation in the \WCR{}.
This approach decouples the charge-asymmetric process, \Wjets{}, from possible mis-modelling of the \ttbar{} events, which is a charge-symmetric process.
This method is also useful for separating charge-asymmetric events from the VLQ signal, as it is also charge-symmetric.
The contributions to $A$ from charge-symmetric backgrounds, such as \ttbar{} and multijet, cancel out.
The small contributions from other charge-asymmetric backgrounds, such as single top, diboson, and $t$-associated production, is accounted for using MC samples.
The dependence of the \Wjets{} normalisation on \ST{} and the $W$-tag requirements was investigated in the control region sidebands (relaxed or tightened requirements on \ST{} or the $W$-tag selection in the \WCR{}) to ensure applicability of this correction in the signal region.
The \Wjets{} correction is derived as a function of \ST{} up to $\ST=\SI{4}{\TeV}$ as a check and no dependence of the \Wjets{} correction on the event \ST{} is observed.
An additional uncertainty is added to cover the difference between the normalisation correction found in the \WCR{} and the control region sidebands with varied $W$-tag requirements.
The dependence of the \Wjets{} normalisation on other relevant event and kinematic variables is checked and no significant dependencies are observed.
Thus, a single normalisation factor is applied to the \Wjets{} MC prediction.
The \Wjets{} normalisation correction amounts to $\displaystyle f_{W\mathrm{+jets}} = 0.915 \pm 0.09 \text{ (stat.)} \pm 0.54 \text{ (syst.)}$,
the largest systematic uncertainties are from the modelling of the single top background and from inverting the requirements of the $W$-tag selection.

\begin{figure}[htbp]
\centering
\subfloat[]{
\includegraphics[width=0.49\textwidth]{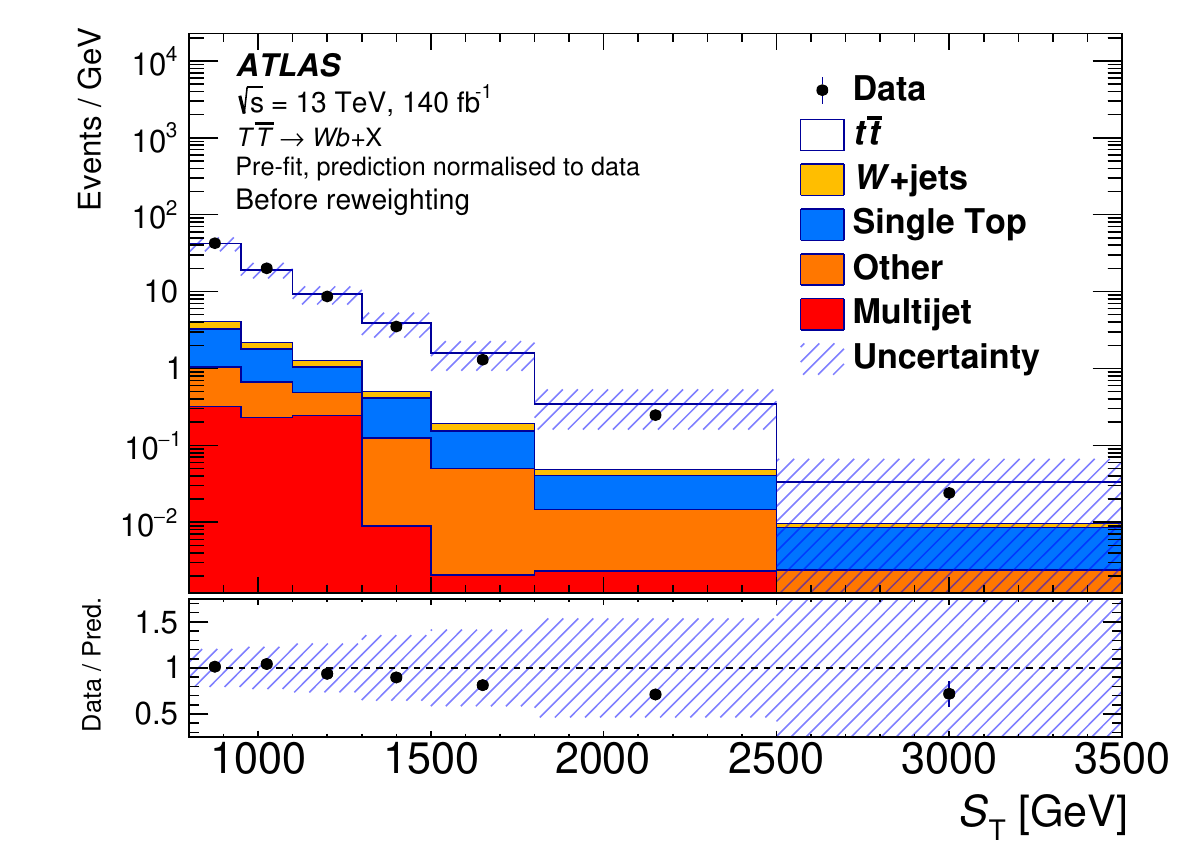}
\label{fig:ttCR2RW_a}
}
\subfloat[]{
\includegraphics[width=0.49\textwidth]{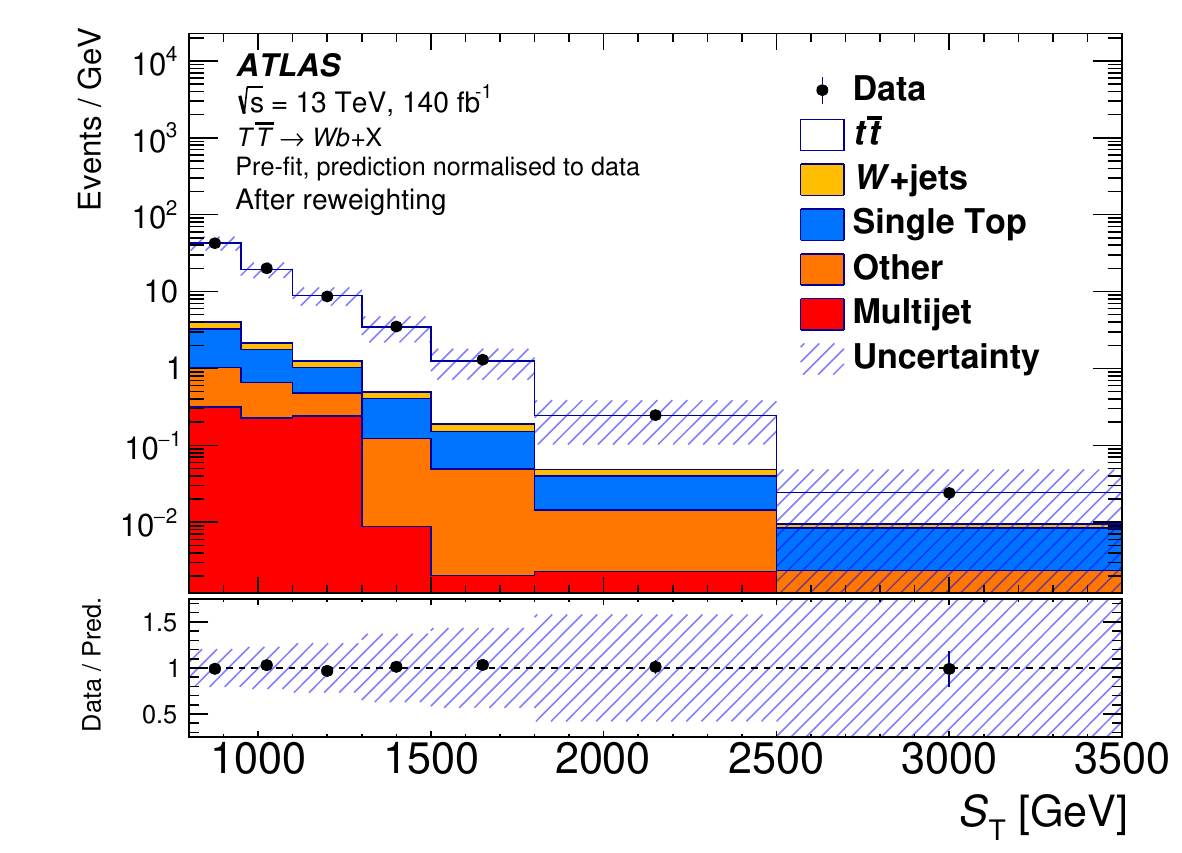}
\label{fig:ttCR2RW_b}
 
}
 
\caption{Distributions of \ST{} for data (dots) and predictions (histograms with various colours) in the \ttRW{} \protect\subref{fig:ttCR2RW_a} before and \protect\subref{fig:ttCR2RW_b} after applying the reweighting (see text). The prediction is normalised to the data to illustrate the discrepancies in the shape of the distribution and the effect of the \ST{} shape reweighting. The error bars include the statistical uncertainty and the shaded band represents the total systematic uncertainty. The last bin includes the overflow.}
\label{fig:ttCR2RW}
\end{figure}

The \ttbar{} background estimate from MC simulation is known to overestimate the number of events at high $t$-quark \pt{}~\cite{TOPQ-2016-04}, directly impacting the modelling of \ST{}, which is essential for discriminating signal from background.
Therefore, a data-driven reweighting is applied to the \ttbar{} MC events to correct for the difference  between the MC and the data
in the \ST{} distribution.
The correction is derived in the \ttbar{} reweighting region \ttRW{}, defined in Table~\ref{tab:Regions}, which has a \ttbar{} purity of 89\%, with the next most significant contribution from single top background at 6\%.
The main selection criteria differentiating the \ttRW{} from the signal region are $\Delta R(\Wtag,\bhad) < 1.0$ and  $\ST{} >\SI{800}{\GeV}$.
These criteria are motivated by the low angular distance between the $W$ boson and $b$-quark from the decay of a high-\pt{} $t$-quark. 
To ensure that the contamination by signal event in \ttRW{} is low, the invariant masses of the reconstructed VLQs, \mTlep{} and \mThad{}, are required to be smaller than $\SI{700}{\GeV}$.
Figure~\ref{fig:ttCR2RW}\protect\subref{fig:ttCR2RW_a} shows the \ST{} distributions in the \ttRW{} before deriving the reweighting.
 
The reweighting is constructed to correct the shape of the \ST{} distribution, but not the normalisation, as it is varying freely in the final likelihood fit and ultimately constrained by the data in the \ttCR{}.
The contributions to the \ttRW{} from processes other than \ttbar{} are subtracted from the data and the ratio of the resulting \ST{} distribution and the \ST{} distribution from the \ttbar{} prediction is calculated.
The correction factor is then derived in a fit to that ratio with a function of the form:
\begin{equation}\label{eq:STFit}
f(\ST) = p_{0} \cdot \exp(-p_{1} \cdot (\ST - p_{2})^{2}) + p_{3},
\end{equation}
where $p_i$ (with $i=0,1,2,3$) are the free parameters of the fit.
Other functional forms are also studied, but the one in Eq.~\eqref{eq:STFit} is found to describe the data/MC ratio most accurately.
The fit is also found to be consistent for different fit intervals and bin choices.
The \ST{} and the jet multiplicity, $N_\text{jets}$, are correlated, so the correction to \ST{} could also depend on $N_\text{jets}$.
Therefore, separate fits are performed for events with $3 \leq N_\text{jets} \leq 6$ and $N_\text{jets} \geq 7$.
Figure~\ref{fig:ttCR2RW}\protect\subref{fig:ttCR2RW_b} shows the \ST{} distribution in the \ttRW{} after the application of the reweighting correction.
 
The statistical uncertainty of the fit is propagated as an uncertainty on the reweighting correction.
The impact of the uncertainty in the single top modelling is taken into account by varying the single top contribution by 100\% in the \ttRW{} region.
Uncertainties in other processes have negligible impact on the extracted reweighting.
The offset parameter $p_{3}$ in Eq.~\eqref{eq:STFit} causes the function to approach an asymptote at high values of \ST{} to 0.64 for $3 \leq N_\text{jets} \leq 6$ and 0.62 for $N_\text{jets} \geq 7$, where the statistical uncertainty is too large to determine the exact dependence of the correction on $S_{T}$ with confidence.
In addition, an uncertainty of 58\% (38\%), covering the large statistical uncertainty in the high-\ST{} tail, is added if $\ST{} \geq \SI{2.5}{\TeV}$ for events with $3 \leq N_\text{jets} \leq 6$ ($N_\text{jets} \geq 7$).
Finally, the correction in \ST{} is tested in a dedicated \ttbar{}-dominated validation region, where the \ttbar{} MC estimate is found to have good agreement with data after application of the \ST{}-dependent correction.
 
Events from multijet production satisfy the lepton requirement of the signal region by mis-identifying  a hadronic jet as a lepton or a non-prompt lepton from a heavy-flavour jet, collectively referred to as non-prompt leptons.
The contribution from multijet production is small, accounting for about 5\% of the events in the signal region.
Background from these processes is estimated by using a data-driven``Matrix Method'' (MM)~\cite{EGAM-2019-01}.
This method uses the classification of the leptons in ``loose'' and ``tight'' categories to predict the number of non-prompt leptons.
Tight leptons correspond to signal leptons, as described in Section~\ref{sec:event}.  Loose leptons satisfy the baseline criteria, as defined in Section~\ref{sec:event}, but fail to meet the tight lepton requirements.
A ``loose region''  enriched in non-prompt leptons is created by changing the lepton requirement from tight to loose for the signal region selection.
The number of events with prompt (non-prompt) leptons in the signal region is related to the number of events with prompt (non-prompt) leptons in the loose region by the efficiencies for prompt (non-prompt) leptons that satisfy the baseline lepton requirements to also satisfy the tight requirements, which is referred to as the ``real efficiency'' (``fake efficiency'').
Therefore, once the real and fake efficiencies are determined, the number of events with non-prompt leptons (the multijet events) in the signal region can be calculated, as described in Ref.~\cite{ATLAS-CONF-2014-058}.
 
The tight lepton criteria reduces the contribution of the multijet background in the signal and control regions to very low levels, such that the shapes of differential distributions estimated by the MM are dominated by statistical fluctuations.
An accurate prediction for the multijet background is particularly important for the \mTlep{} distribution, as it is used in the final likelihood fit.
As the background events do not contain real VLQ decays, the distribution of reconstructed masses is largely determined by the kinematic constraints from the event selection.
As a result, the \mTlep{} distribution for \Zjets{} and multijet events are very similar, but is much more stable for \Zjets{}.
Therefore, in the final fit, the shape of the \mTlep{} template for the multijet background is taken from \Zjets{} events, while the total predicted yield is determined from the MM calculation.
 
For reconstructed \mTlep{} below \SI{200}{\GeV}, the MM estimate for multijet events is extremely sensitive to the amount of \ttbar{} that is subtracted from the data.
Therefore, an iterative method is employed to determine the total predicted multijet yield for the scaling of the Z+jets template.
First, the MM prediction is evaluated using the nominal normalization for the subtracted \ttbar{}.
Second, the \mTlep{} template (taken from \Zjets{} events) is scaled to have the same yield as the MM prediction for events with $\mTlep{} > \SI{200}{\GeV}$.
Third, the normalisation of the \ttbar{} events is adjusted to get the MM prediction to agree with the \mTlep{} template for $\mTlep{} > \SI{200}{\GeV}$.
The second and third steps are then repeated until the predicted multijet yield changes by less than 1\% from the previous iteration.
An uncertainty of 100\% is assigned to the multijet yield to account for the impact of systematic uncertainties in the modelling of prompt lepton processes and the fake-lepton efficiency.
An additional uncorrelated uncertainty of 100\% is applied to the multijet prediction for $\mTlep{} < \SI{200}{\GeV}$ to account for the uncertainty of the MM prediction in that region due to the normalization of the \ttbar{} background.
To model the multijet shape in other variables and analysis regions, the MC sample for multijet production, scaled to the MM estimate, is used.


\section{Systematic uncertainties}
\label{sec:systematics}

 
Systematic uncertainties are broadly grouped into the following two classes:
experimental uncertainties, which are related to the modelling of the detector response and reconstruction  of physics objects, and to the data-driven background estimate or the data-driven corrections to the simulated backgrounds;
and theoretical uncertainties, which are related to the modelling of the physics processes by simulation.
Unless stated otherwise, uncertainties from a common source are correlated across processes and regions in the final statistical analysis and the impact of the uncertainties is allowed to be constrained by the likelihood fit to the data.

The largest systematic uncertainties  are  related to the modelling of  \ttbar{} and single top backgrounds.
The uncertainties in the theory for the \ttbar{} prediction include the choice of MC generator (in this case, MC events were generated to NLO accuracy); ISR, FSR, and parton shower model; and shower matching scheme of simulated \ttbar{} events.  Each of these uncertainties is evaluated by comparing the nominal \ttbar{} prediction to the alternative MC samples, as described in Section~\ref{sec:samples}.
Modelling uncertainties in the single top prediction include the choice of the parton shower and the matching of the NLO matrix element to the parton shower, which are evaluated using alternative samples as described in Section~\ref{sec:samples}.
Similarly, an uncertainty due to the choice of the $\ttbar/Wt$ overlap removal scheme is evaluated by comparing the nominal single top MC produced with DR scheme to the alternative sample produced with the DS scheme~\cite{ATL-PHYS-PUB-2021-042}.
The DR scheme is found to be more compatible with the data at lower energies, while at higher energies, the DS scheme is observed to provide a better description of the data.
To avoid extrapolation of the DS versus DR uncertainty from the regions defined with low \ST{} to those at high \ST{}, this uncertainty is split into a high-\ST{} component, applied to regions with $\ST{}>\SI{1.9}{\TeV}$, and a low-\ST{} component, applied to regions with $1.4 < \ST{} < \SI{1.9}{\TeV}$.
For both the single top and the \ttbar{} production, uncertainties in the PDF are obtained using the \texttt{PDF4LHC15} combined PDF set~\cite{Butterworth_2016}.
The effect of QCD scale uncertainties is estimated by independently doubling or halving the renormalisation and factorisation scales for single top and \ttbar{} production, respectively.
The QCD scale variation with the largest impact on the fit discriminant (\mTlep{}) is used.

To simplify the statistical analysis, all the smaller backgrounds from \ttbar{}$V$, \Zjets{}, rare top processes and diboson production are grouped into a single category named ``other'', and a single overall normalisation uncertainty of 50\% is applied.
This uncertainty of 50\% is adopted from the uncertainty in \Zjets{} events which has the largest relative uncertainty among the processes contained in the ``other'' category.
The uncertainty in the \Zjets{} contribution was estimated in the previous analysis~\cite{EXOT-2016-14} by investigating the mis-modelling of the jet multiplicity of \Zjets{} events.

Experimental uncertainties include an uncertainty of 0.83\% on the integrated luminosity measurement on the combined 2015$-$2018 data~\cite{arXiv:2212.09379}, obtained using the LUCID-2 detector~\cite{Avoni:2633501} for the primary luminosity measurements. 
Uncertainties in leptons arise from potential mis-modelling of the electron and muon energy scales and resolutions~\cite{EGAM-2018-01,MUON-2022-01} and
from uncertainties in the correction factors to the electron and muon trigger, reconstruction, identification and isolation efficiencies~\cite{MUON-2018-03,EGAM-2018-01}.
 
Small-$R$ jets have uncertainties in the JES, the JER~\cite{JETM-2018-05} and on the JMS~\cite{JETM-2018-02}.
An uncertainty is assigned to the JVT selection efficiency~\cite{PERF-2014-03} and to the reweighting factors that correct the pile-up profile in MC simulations to match that in data.
Similarly, JES, JER~\cite{JETM-2018-02} and JMS uncertainties, and in addition JMR~\cite{ATLAS-CONF-2020-022} systematic uncertainties
are assigned to large-radius jets.
An uncertainty related to the scale and resolution of the track soft term in the \MET{} calculation~\cite{PERF-2016-07} is applied.
 
Uncertainties in the $b$-tagging algorithm selection efficiency include uncertainties in the $b$-jet selection efficiency, $c$-jet and light jet mistag rate
correction factors and additional components from the extrapolation of the calibrations to high \pT{} and from $c$-jets to $\tau$ leptons~\cite{FTAG-2018-01,FTAG-2020-08,arXiv:2301.06319}.
An uncertainty is assigned to the efficiency correction of the $W$ boson tagging algorithm~\cite{ATL-PHYS-PUB-2020-017,JETM-2018-03}.

Uncertainties in the data-driven corrections for simulated backgrounds or in the data-driven multijet event estimate are briefly listed below and detailed in Section~\ref{sec:bkg}.
The uncertainty in the data-driven correction for simulated $W$+jets events is 60\%.
This is among the dominant systematic uncertainties, though still much smaller than the statistical uncertainty of the data set.
The uncertainty in the \ttbar{} \ST{} shape reweighting consists of a statistical component and a systematic component from the uncertainty in the single top contribution in the \ttRW{}.
The two components affecting the multijet background estimate include a global 100\% uncertainty in the MM estimate and an additional 100\% uncertainty for reconstructed VLQ invariant masses \mTlep{} below 200~\GeV.


\section{Statistical analysis and results}
\label{sec:result}


The presence of the VLQ signal is tested by using a fit  to the reconstructed mass distributions of the leptonically decaying $T$ candidate (\mTlep{}) from the signal regions SR1 and SR2 and control regions \ttCR{}, \lowSTCR{}, and \highSTCR{}, denoted fit regions.
The mass of the leptonically decaying $T$-quark is chosen because it is found to have a better resolution and sensitivity than the mass of the hadronically decaying $T$-quark.
The fit maximises a binned likelihood function, $\mathcal{L}(\mu,\boldsymbol{\theta})$,
constructed as a product of Poisson probabilities for all bins considered in the search and depends on the parameter of interest $\mu$ and a vector of nuisance parameters (NP) $\boldsymbol{\theta}$.
The parameter of interest is the signal strength $\mu = \sigma^\text{test}/\sigma^\text{theory}$, where $\sigma^\text{test}$ is the value for the VLQ cross-section being tested and $\sigma^\text{theory}$ is the theoretical prediction.
Each NP ${\theta_i}$ encodes a systematic uncertainty with a Gaussian function prior, except the normalisations of the \ttbar{} and single top backgrounds, which are unconstrained, and the statistical uncertainties due to the finite size of the MC samples, which are included as one additional NP per bin that accounts for the statistical uncertainty of the total estimated yield in the given bin.
If an uncertainty would impact the normalisation or shape of all bins for a given process by less than 1\%, the NP is removed from the fit for that process.
The test statistic $q_{\mu}$ is defined as the profile likelihood ratio,
$q_{\mu}=-2\mathrm{ln}(\mathcal{L}(\mu,\hat{\hat{\boldsymbol{\theta}}}_{\mu})/ \mathcal{L}(\hat{\mu}, \hat{\boldsymbol{\theta}}))$,
where $\hat{\mu}$ and $\hat{\boldsymbol{\theta}}$ are the values of the parameters that maximise the likelihood function (with the constraint 0$\leq  \hat{\mu} \leq \mu$), and $\hat{\hat{\boldsymbol{\theta}}}_{\mu}$ are the values of the nuisance parameters that maximise the likelihood function for a given value of $\mu$.
The compatibility  with the background-only hypothesis for the observed data is tested by setting $\mu=0$ in the profile likelihood ratio.
Upper limits on the cross-section times branching ratio for a given signal are derived by using $q_{\mu}$ in the modified frequentist CL$_{\mathrm{s}}$ method~\cite{cls_2, cls}, approximated using the asymptotic formulae~\cite{cls_3}.
For a given signal, the cross-section times branching ratio is excluded at $\geq$ 95\% confidence level (CL) when CL$_{\mathrm{s}}<0.05$.

\begin{figure}[htb]
\centering
\subfloat[]{
\includegraphics[width=0.32\textwidth]{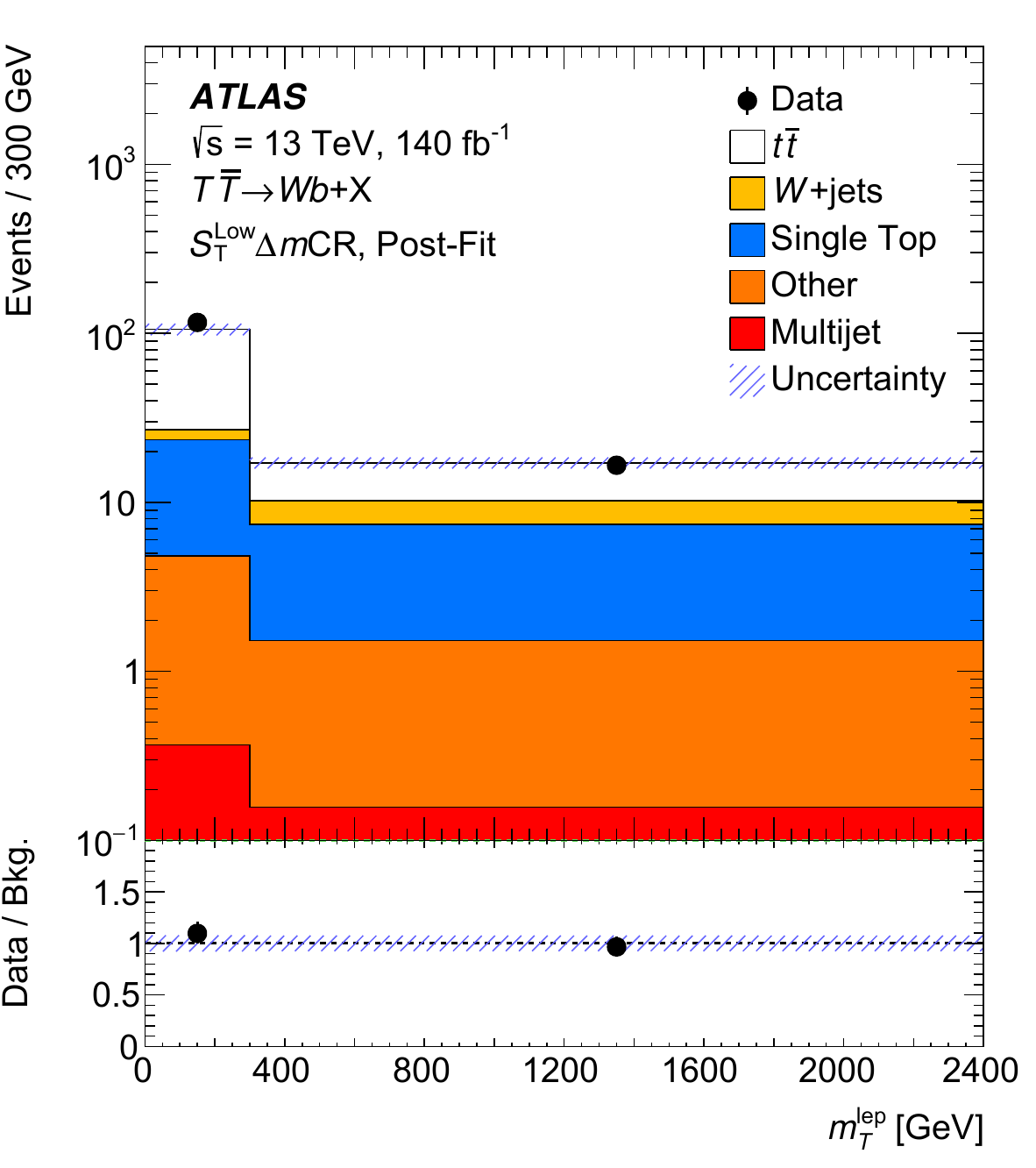}
\label{fig:PostFit_a}
}
\subfloat[]{
\includegraphics[width=0.32\textwidth]{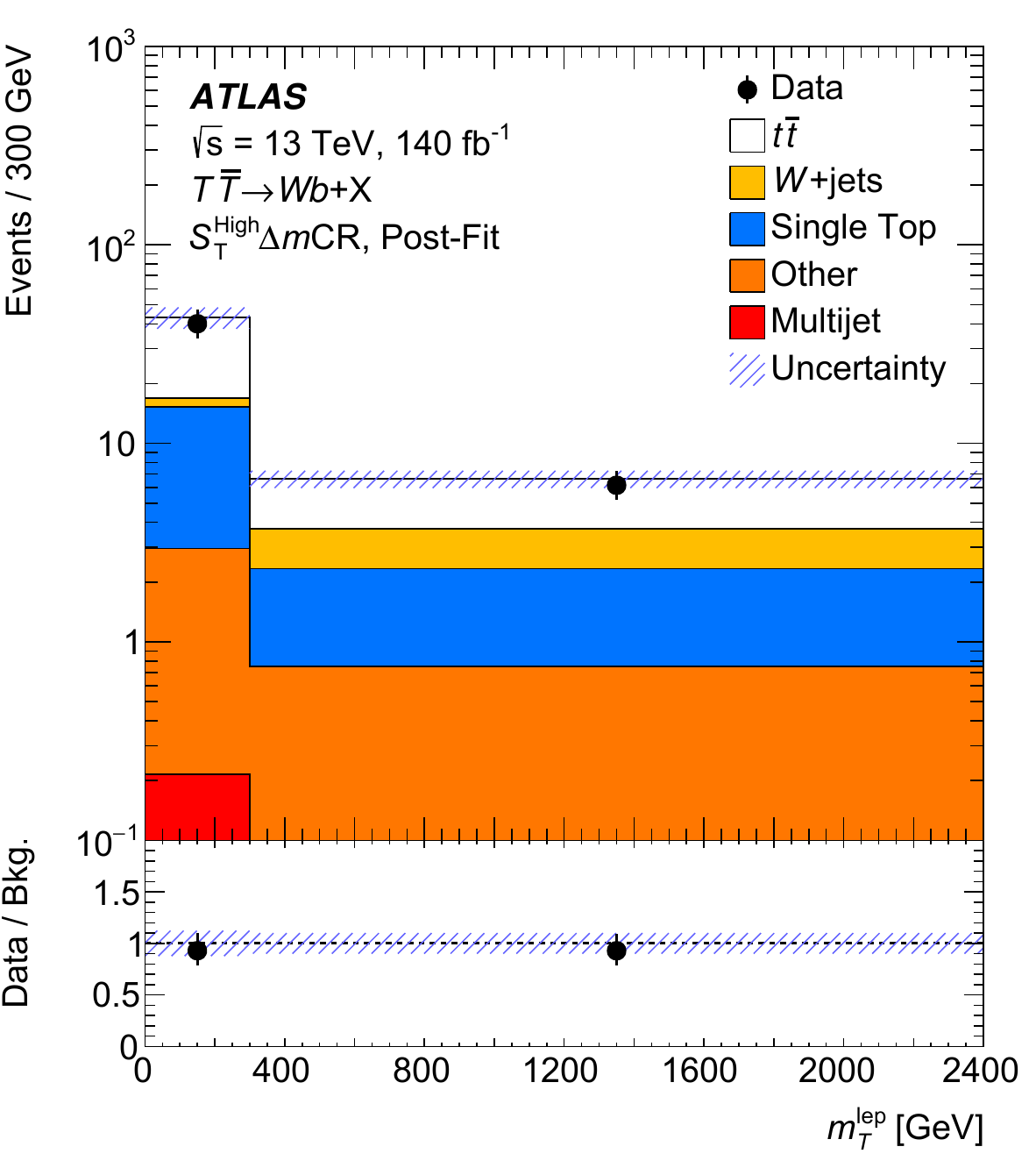} \label{fig:PostFit_b}
}\\
\subfloat[]{
\includegraphics[width=0.32\textwidth]{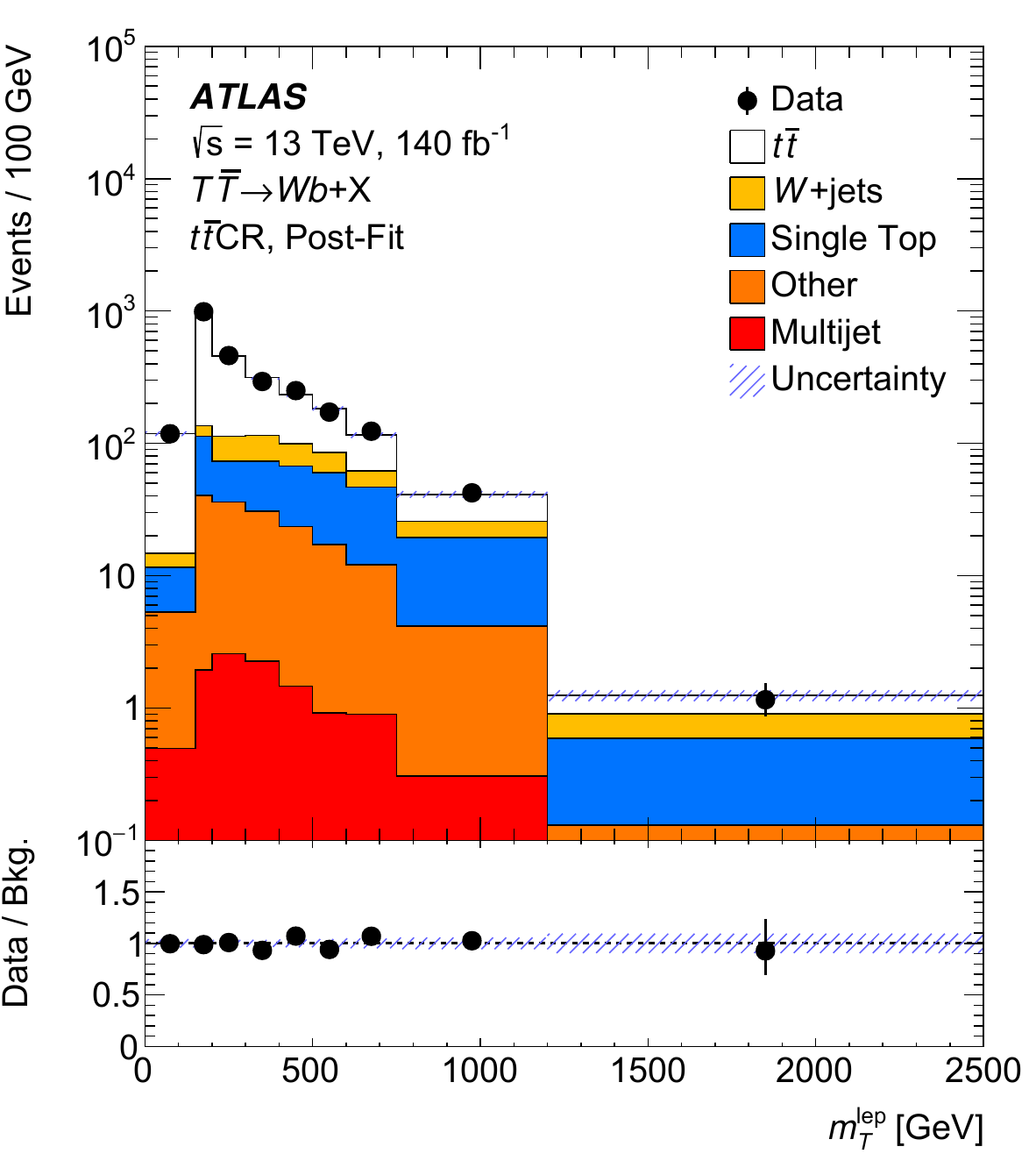}
\label{fig:PostFit_c}
}
\subfloat[]{
\includegraphics[width=0.32\textwidth]{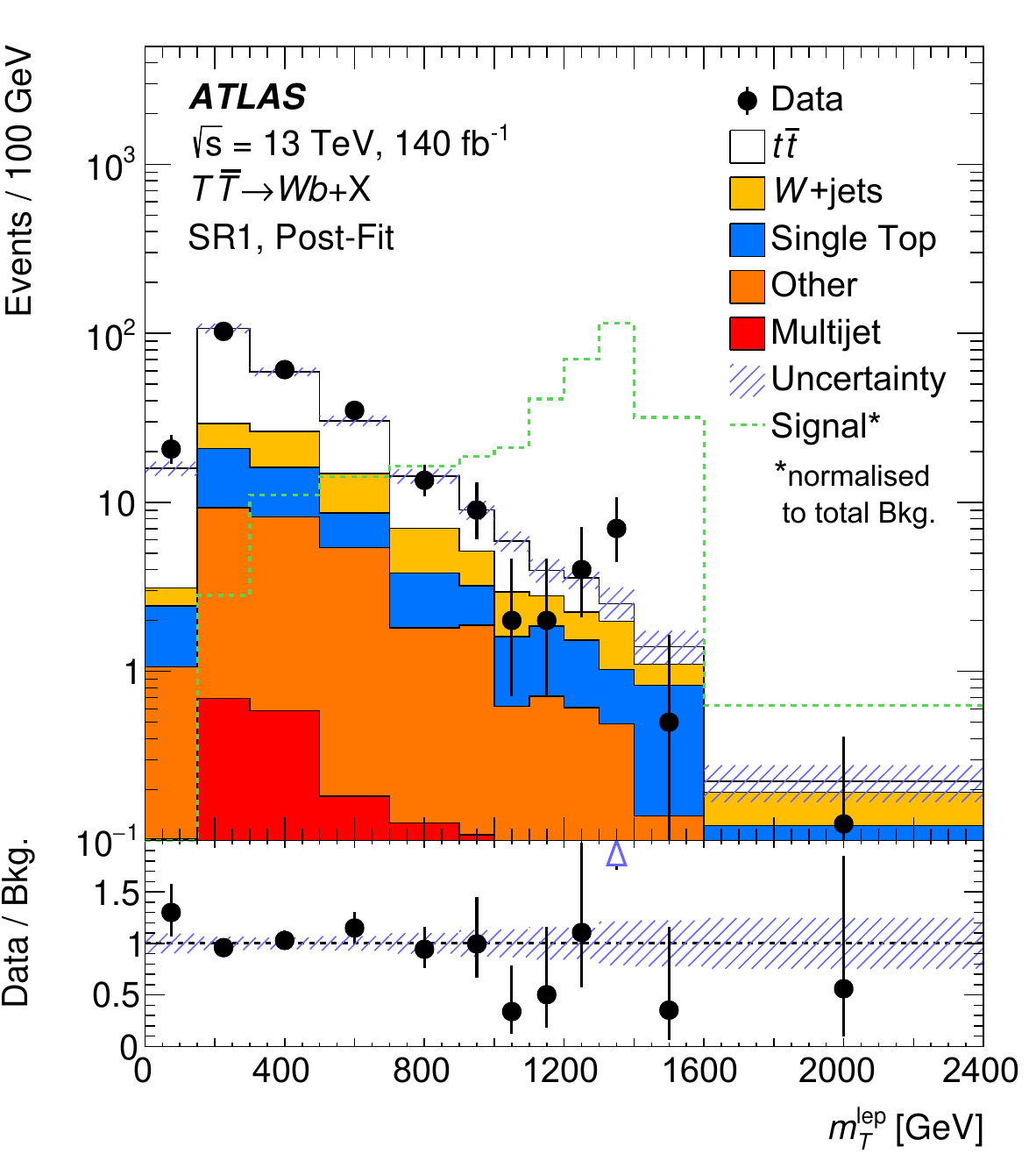}
\label{fig:PostFit_d}
}
\subfloat[]{
\includegraphics[width=0.32\textwidth]{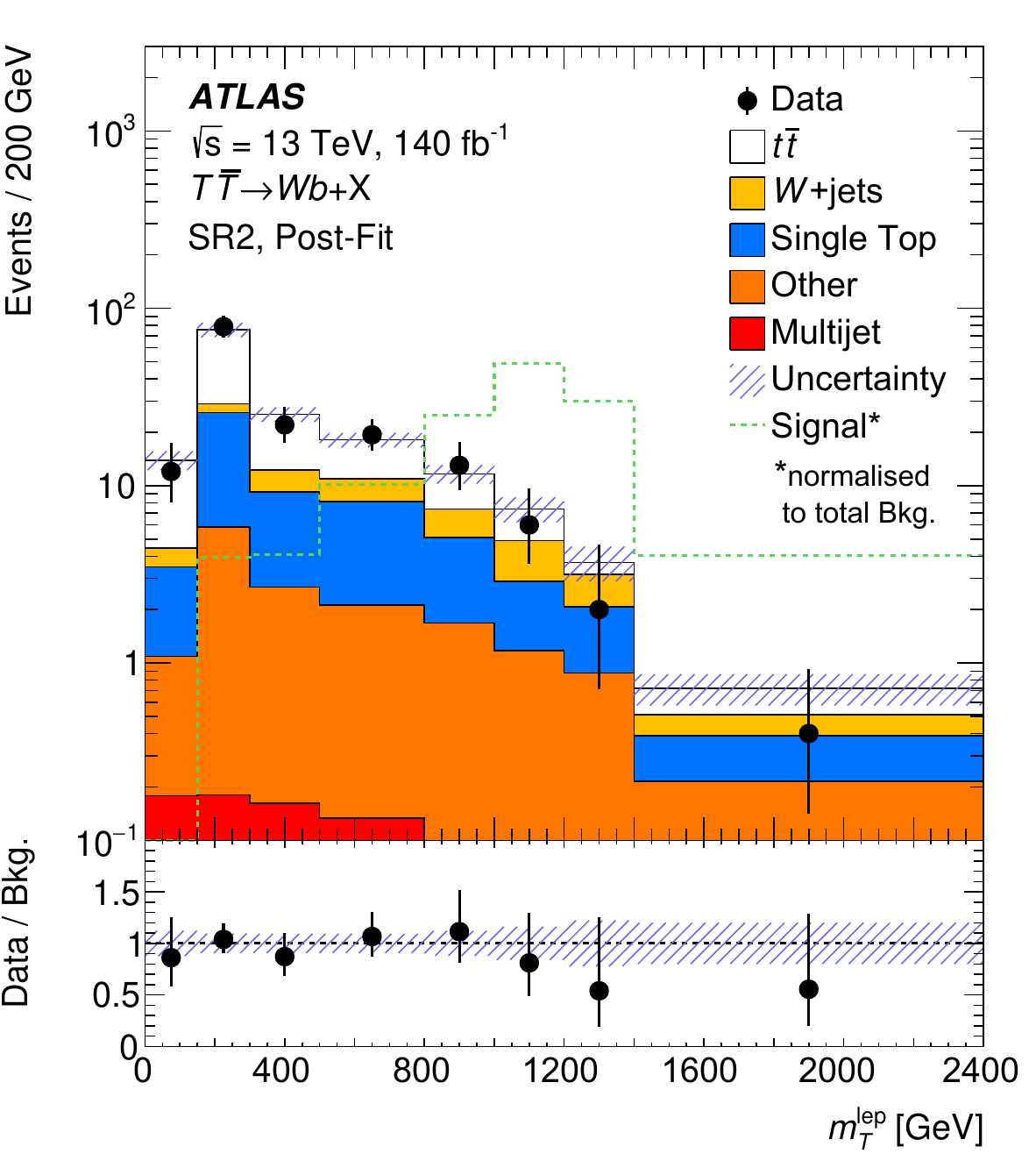}
\label{fig:PostFit_e}
}
\hfill
\caption{Distribution for \mTlep{} for data (dots) and predictions (histograms with various colors) in the five fit regions (\protect\subref{fig:PostFit_a} \lowSTCR{}, \protect\subref{fig:PostFit_b} \highSTCR, \protect\subref{fig:PostFit_c} \ttCR{}, \protect\subref{fig:PostFit_d} SR1, and \protect\subref{fig:PostFit_e} SR2, respectively) after the simultaneous fit to data under the background-only hypothesis (Post-Fit). The binning in \mTlep{} was optimised to maximise the sensitivity to the signal in the signal regions while it is ensured that each bin of the fit regions contains a reasonable number of data events to guarantee good fit behaviour. The lower panels of each plot shows the ratio of data and predicted background yields. The arrow indicates that the ratio point is outside the ordinate range. The overflow events are included in the last bin. The error bars include the statistical uncertainty and the shaded band represents the systematic uncertainty after the likelihood fit (see text). The dashed green histograms in \protect\subref{fig:PostFit_d} SR1 and \protect\subref{fig:PostFit_e} SR2 show the shape for the signal with $\mT = \SI{1400}{\GeV}$ and $\BR{T \to Wb} = 1$ scaled to the total background prediction.  The expected signal event yields in each of the fit regions are displayed in Table~\ref{tab:VLQYields}.
}
\label{fig:PostFit}
\end{figure}

Figure~\ref{fig:PostFit} shows the \mTlep{} distribution in each of the five fit regions after the simultaneous fit to the background-only hypothesis to data.
The \mTlep{} distributions of the SM backgrounds show a steep decline towards high \mTlep{}, whereas the signal \mTlep{} distribution is expected to peak close to the simulated \mT{}, at high \mTlep{} in \SRa{} and \SRb{}, effectively distinguishing it from the SM backgrounds.
The corresponding event yields are listed in Table~\ref{tab:FinalYields}.
The uncertainty in the total prediction does not equal the sum in quadrature of the individual component due to correlations between the fit parameters.
The expected number of \TTbar{} signal events in each of the five fit regions are given in Table~\ref{tab:VLQYields} for various VLQ scenarios.
The compatibility with the background-only hypothesis  for the data  is estimated by integrating the distribution of the test statistic above the observed value of $q_{0}$.
This value is computed for each signal scenario considered, defined by the assumed mass of the $T$-quark and the three decay branching ratios.
These results assume the $T$-quark has a narrow width.

\begin{table}[htb]
\centering
\caption{Event yields in the five fit regions after the fit to the data under the background-only hypothesis.  The uncertainties include statistical and systematic uncertainties. The uncertainties in the individual background components can be larger than the uncertainty in the sum of the backgrounds due to correlations. }
\label{tab:FinalYields}
\begin{tabular}{l
S[table-format=5.0(4)] 
S[table-format=4.1(4)]
S[table-format=4.1(4)]
S[table-format=6.1(4)]
S[table-format=5.0(4)]}
\hline
& {SR1} & {SR2} & {\highSTCR{}} & {\ttCR{}} & {\lowSTCR{}} \\
\hline
\ttbar{}   & 257 \pm 33 & 74 \pm 14 & 46 \pm 9 & 1513 \pm 150 & 127 \pm 18 \\
\Wjets{}   & 61 \pm 25 & 17 \pm 7 & 12 \pm 5 & 213 \pm 90 & 24 \pm 10 \\
Single Top   & 53 \pm 19 & 40 \pm 14 & 23 \pm 10 & 340 \pm 120 & 60 \pm 20 \\
Other   & 48 \pm 21 & 15 \pm 7 & 7.5 \pm 3.4 & 160 \pm 70 & 14 \pm 6 \\
Multijet   & 3 \pm 11 & 0.9 \pm 3.1 & 0.7 \pm 2.4 & 10 \pm 40 & 1 \pm 5 \\
\hline
Total  & 422 \pm 21 & 146 \pm 11 & 89 \pm 8 & 2240 \pm 60 & 226 \pm 13 \\
\hline
Data   & 430 & 142 & 83 & 2235 & 232 \\
\hline
\end{tabular}
\end{table}
 
\begin{figure}[htb]
\centering
\subfloat[]{
\includegraphics[width=0.49\textwidth]{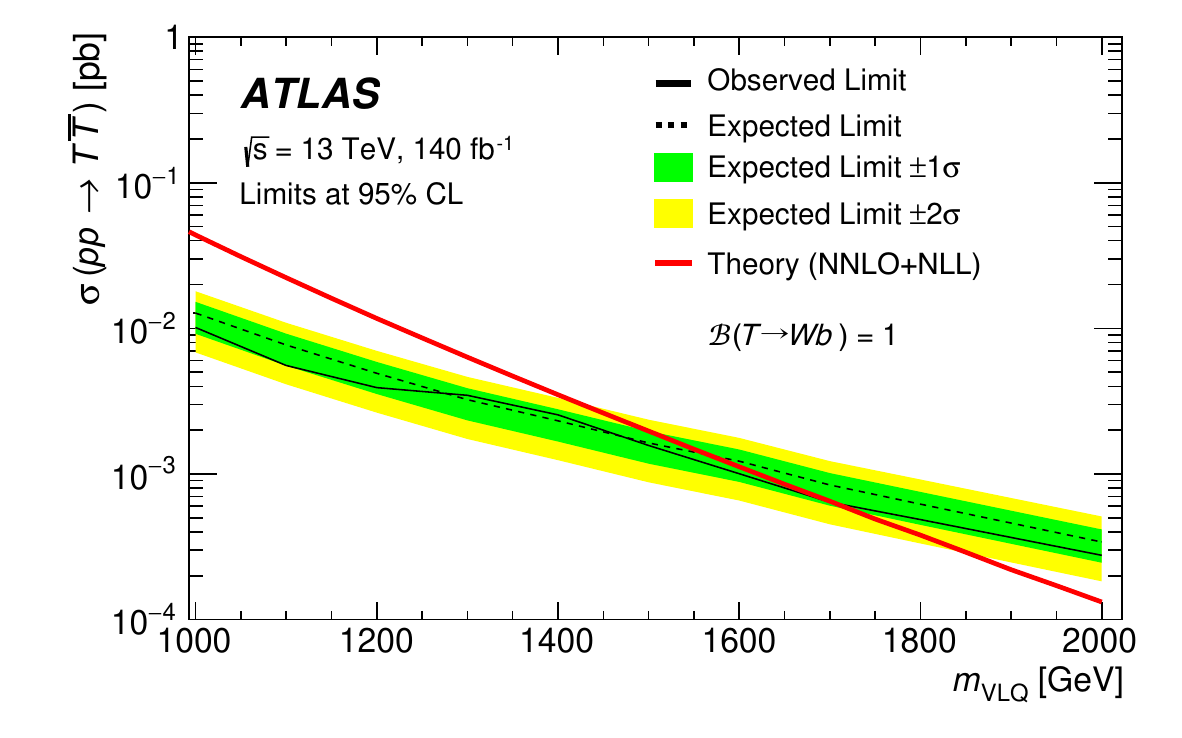}
\label{fig:LimitXS_a}
}
\subfloat[]{
\includegraphics[width=0.49\textwidth]{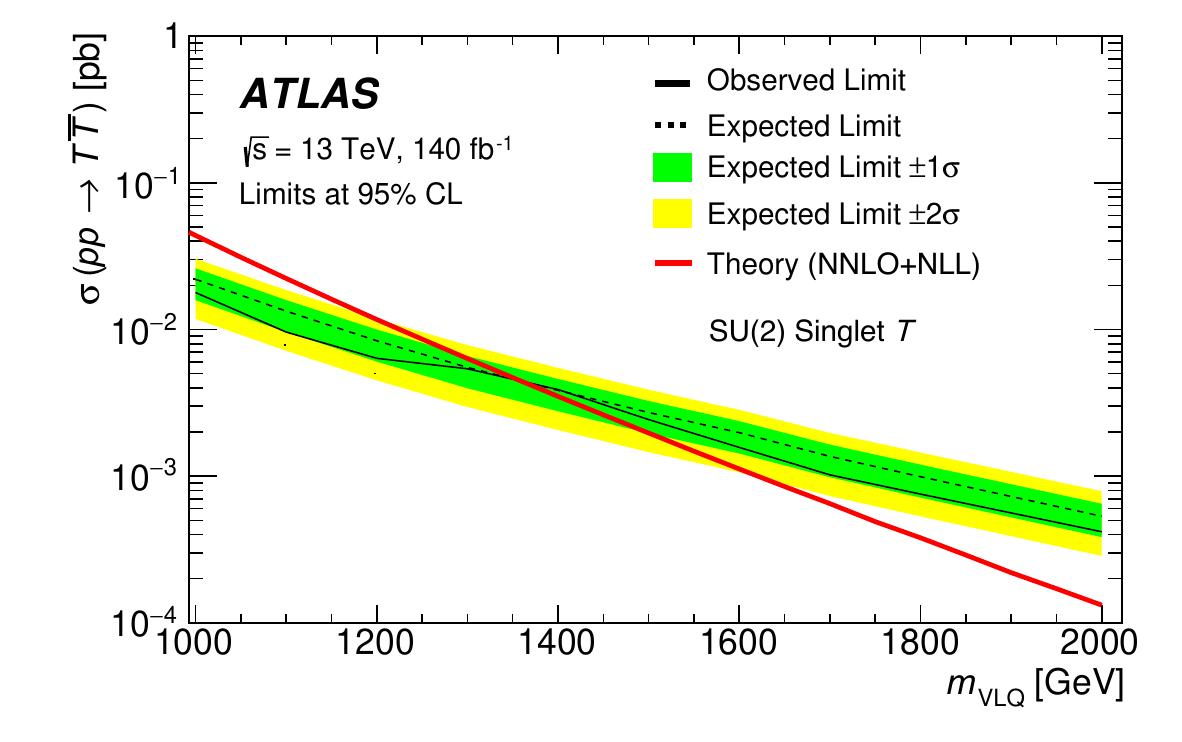}
\label{fig:LimitXS_b}
}
\caption{Expected (dashed black line) and observed (solid black line) upper limits at the 95\% CL on the $T\bar{T}$ cross-section as a function of $T$-quark mass for \protect\subref{fig:LimitXS_a} the $\BR{T \to Wb} = 1$ scenario and \protect\subref{fig:LimitXS_b} in the SU(2) singlet $T$ scenario. The green and yellow bands correspond to $\pm$1 and $\pm$2 standard deviations around the expected limit, respectively. The thin red line shows the theoretical prediction.}
\label{fig:LimitXS}
\end{figure}

\begin{table}[htb]
\centering
\caption{Expected \TTbar{} event yields in the five fit regions for various VLQ scenarios.  The uncertainties include statistical and systematic uncertainties.}
\label{tab:VLQYields}
\begin{tabular}{c
S[table-format=2.2(2)]
S[table-format=2.2(2)]
S[table-format=1.2(2)]
S[table-format=1.4(4)]
S[table-format=1.4(4)]
}
\hline
\multicolumn{3}{l}{Yields for $\BR{T \to Wb}=1 :$} \\
\hline
$\mT/\GeV$  & {SR1} & {SR2} & {\highSTCR{}} & {\ttCR{}} & {\lowSTCR{}} \\
{1200} & 83 \pm 4 & 26.9 \pm 1.3 & 9.5 \pm 0.6 & 18.7 \pm 0.6 & 1.65 \pm 0.11 \\
{1400} & 27.6 \pm 1.7 & 10.6 \pm 0.7 & 4.6 \pm 0.4 & 2.41 \pm 0.10 & 0.24 \pm 0.04 \\
{1600} & 8.6 \pm 0.6 & 3.59 \pm 0.24 & 1.87 \pm 0.14 & 0.273 \pm 0.017 & 0.066 \pm 0.009 \\
{1800} & 2.69 \pm 0.23 & 1.47 \pm 0.11 & 0.81 \pm 0.06 & 0.057 \pm 0.006 & 0.0076 \pm 0.0015 \\
 
\hline
\hline
\multicolumn{3}{l}{Yields for SU(2) singlet $T$ :} \\
\hline
$\mT/\GeV$  & {SR1} & {SR2} & {\highSTCR{}} & {\ttCR{}} & {\lowSTCR{}} \\
{1200} & 42.9 \pm 2.1 & 20.4 \pm 1.2 & 9.8 \pm 0.8 & 12.5 \pm 0.5 & 1.49 \pm 0.08 \\
{1400} & 14.8 \pm 0.9 & 7.7 \pm 0.5 & 4.2 \pm 0.4 & 1.71 \pm 0.08 & 0.267 \pm 0.024 \\
{1600} & 4.66 \pm 0.32 & 2.58 \pm 0.18 & 1.82 \pm 0.16 & 0.246 \pm 0.013 & 0.057 \pm 0.004 \\
{1800} & 1.53 \pm 0.13 & 0.94 \pm 0.07 & 0.75 \pm 0.06 & 0.0526 \pm 0.0023 & 0.0101 \pm 0.0007 \\
\hline
\end{tabular}
\end{table}
No significant excess above the background expectation is found.
Upper limits at the 95\% CL on the $T\bar{T}$ production cross-section
are set for two benchmark scenarios as a function of $T$-quark
mass \mT{} and compared with the theoretical prediction from \TOPpp[2.0]
(see Figure~\ref{fig:LimitXS}).
The resulting lower limit on \mT{} is
determined using the central value of the theoretical cross-section prediction.
 
For $\BR{T \to Wb} = 1$ and $T$-quark masses from \SI{1000}{\GeV} to \SI{2000}{\GeV}, \TTbar{} production cross-sections greater than \SI{10.1}{fb} to \SI{0.27}{fb} are excluded at 95\% CL.
Comparing to the theory cross-section, this results in an observed (expected) lower limit on the $T$-quark mass of $\mT>\MassLimitObsWb{}$ (\MassLimitExpWb{}) for this scenario.
For branching ratios corresponding to the SU(2) singlet $T$ scenario, the observed (expected) 95\% CL lower mass limit is $\mT>\MassLimitObsSinglet{}$ (\MassLimitExpSinglet{}).
The sensitivity of the analysis is limited by the statistical uncertainty in the data sample size.
Including all systematic uncertainties for the $\BR{T \to Wb} = 1$ scenario degrades the expected cross-section limit by less than 3.7\% for any of the tested masses, and reduces the expected mass limit by only \SI{10}{\GeV}.

In the signal region, the prediction is found to slightly overestimate the data in the signal-sensitive high mass tail of the \mTlep{} distribution.
Therefore, the observed limits on the signal cross-section, and thus also on the signal mass \mT{}, are generally more stringent than the expected limits.
Large bin-to-bin variations between 1000 to 1600~\GeV in the \mTlep{} distribution are observed in \SRa{},
with an excess relative to the prediction in one bin and deficits in three bins.
However, no signal model is compatible with the narrow excess in \SRa{}, as the width of the \mTlep{} distribution for any of the signal models would be much wider than the observed one-bin excess.
Additionally, no such excess or large bin-to-bin variations of the data are observed in \SRb{}.
The narrow excess in \SRa{} does lead to weaker observed cross-section limits in an \mT{} interval between around \SI{1250}{\GeV} to \SI{1500}{\GeV} relative to the rest of the \mT{} spectrum.
 
In the previous search by ATLAS performed on a fraction of this data set with an integrated luminosity of 36\fb~\cite{EXOT-2016-14}, observed (expected) 95\% CL mass limits of 1350~\GeV (1310~\GeV) for the scenario $\BR{T \to Wb}=1$ and 1170~\GeV (1080~\GeV) for a singlet $T$ were found.
The present search thus extends the observed mass limits by 350~\GeV and 190~\GeV for the $\BR{T \to Wb} = 1$ and singlet $T$ case, respectively.
The increase of the mass limits can almost entirely be attributed to the increase in the size of the data sample from 36\fb{} to 140\fb{}, but about 15\% (20\%) of the improvement on the mass limits for the $\BR{T \to Wb} = 1$ (singlet $T$) scenario is expected to come from changes in the analysis strategy, especially from improvements to the $W$ boson tagging and the slicing of the signal region into two regions, \SRa{} and \SRb{} according to $\dm$.

To check that the results do not depend on the weak-isospin of the $T$-quark in the simulated signal events, a sample of \TTbar{} events with a mass of 1.2\,\TeV{} was generated for an SU(2) doublet $T$-quark and compared with the nominal sample of the same mass generated with an SU(2) singlet $T$-quark.
Both the expected number of events and expected excluded cross-section are consistent between the two samples.
Thus the limits obtained are also applicable to VLQ models with non-zero weak-isospin.
As there is no explicit use of charge identification, the $\BR{T \to Wb} = 1$ limits are applicable to the pair-production of vector-like $Y$-quarks of charge $-4/3$, which decay exclusively into $Wb$.
 
\begin{figure}[htb]
\centering
\subfloat[]{
\includegraphics[width=0.49\textwidth]{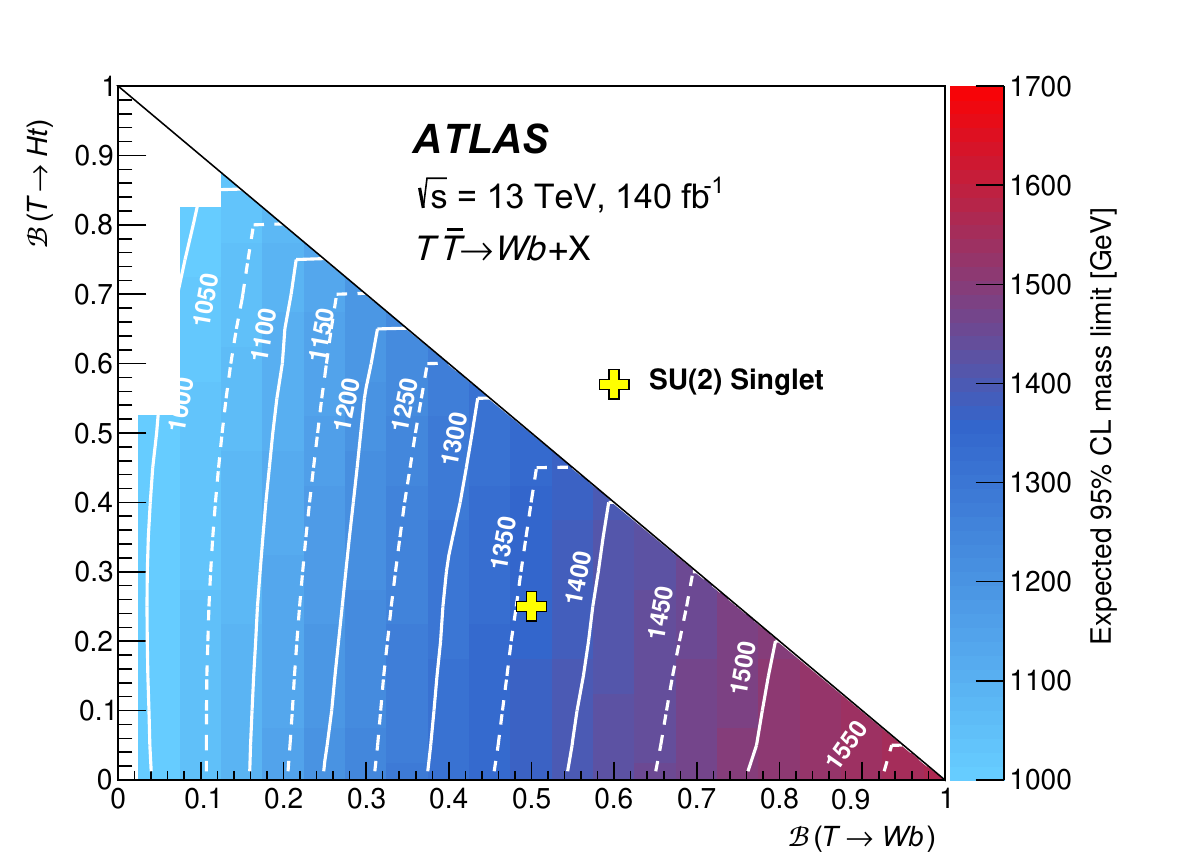}
\label{BRScanT_a}
}
\subfloat[]{
\includegraphics[width=0.49\textwidth]{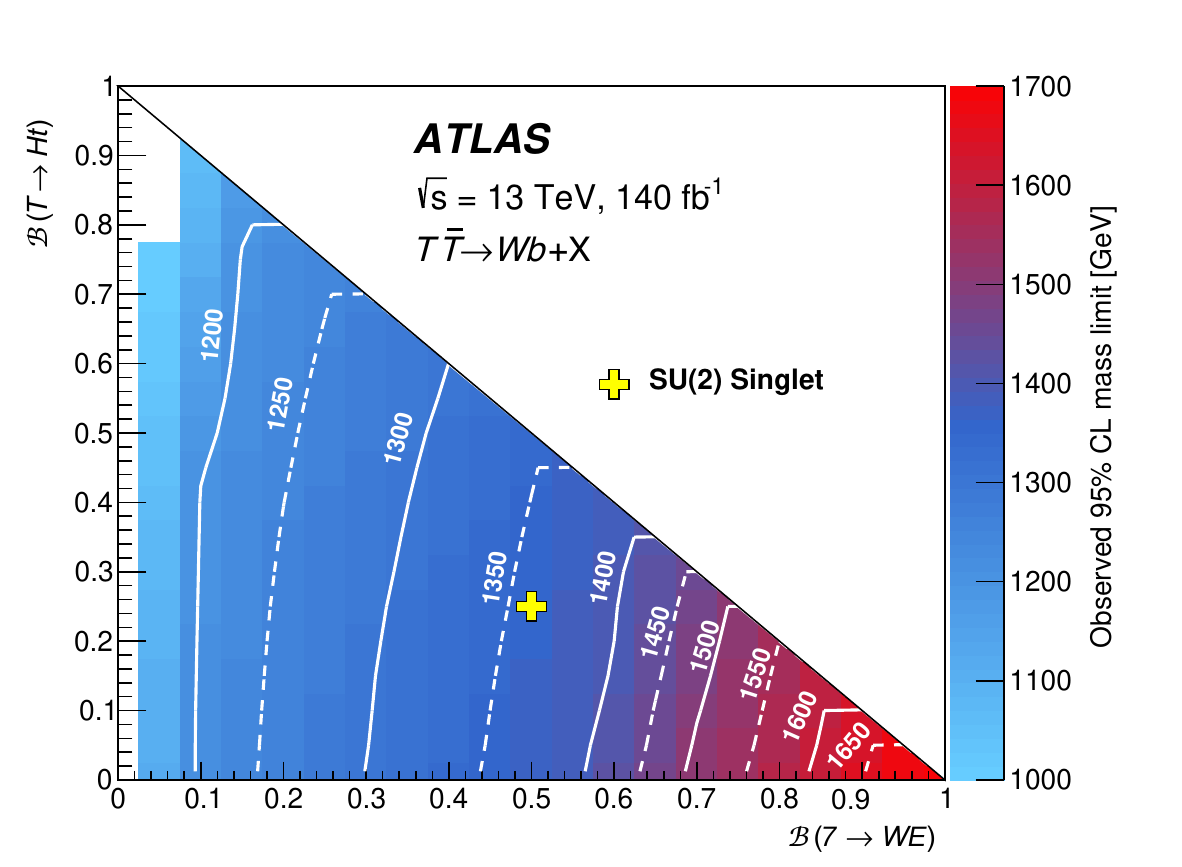}
\label{BRScanT_b}
}
\caption{\protect\subref{BRScanT_a} Expected and \protect\subref{BRScanT_b} observed 95\% CL lower limits on the mass of the $T$-quark in the branching-ratio plane of \BR{T \to Wb} versus \BR{T \to Ht}. Contour lines are provided to guide the eye. The white region is due to the lower mass limit falling below \SI{1000}{\GeV}, the lowest signal mass considered in this search. The yellow marker indicates the branching ratio for the SU(2) singlet $B$ scenario.}
\label{fig:BRScanT}
\end{figure}

\begin{figure}[htb]
\centering
\subfloat[]{
\includegraphics[width=0.49\textwidth]{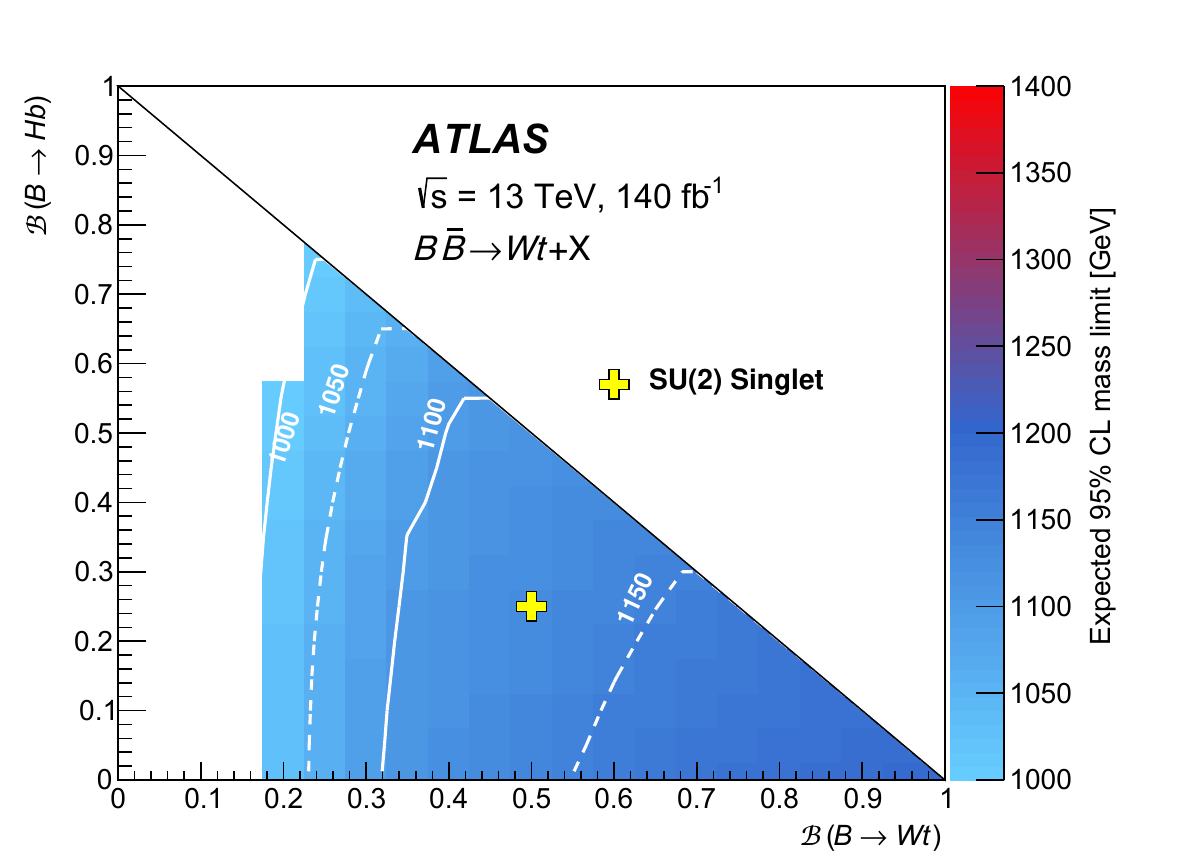}
\label{fig:BRScanB_a}
}
\subfloat[]{
\includegraphics[width=0.49\textwidth]{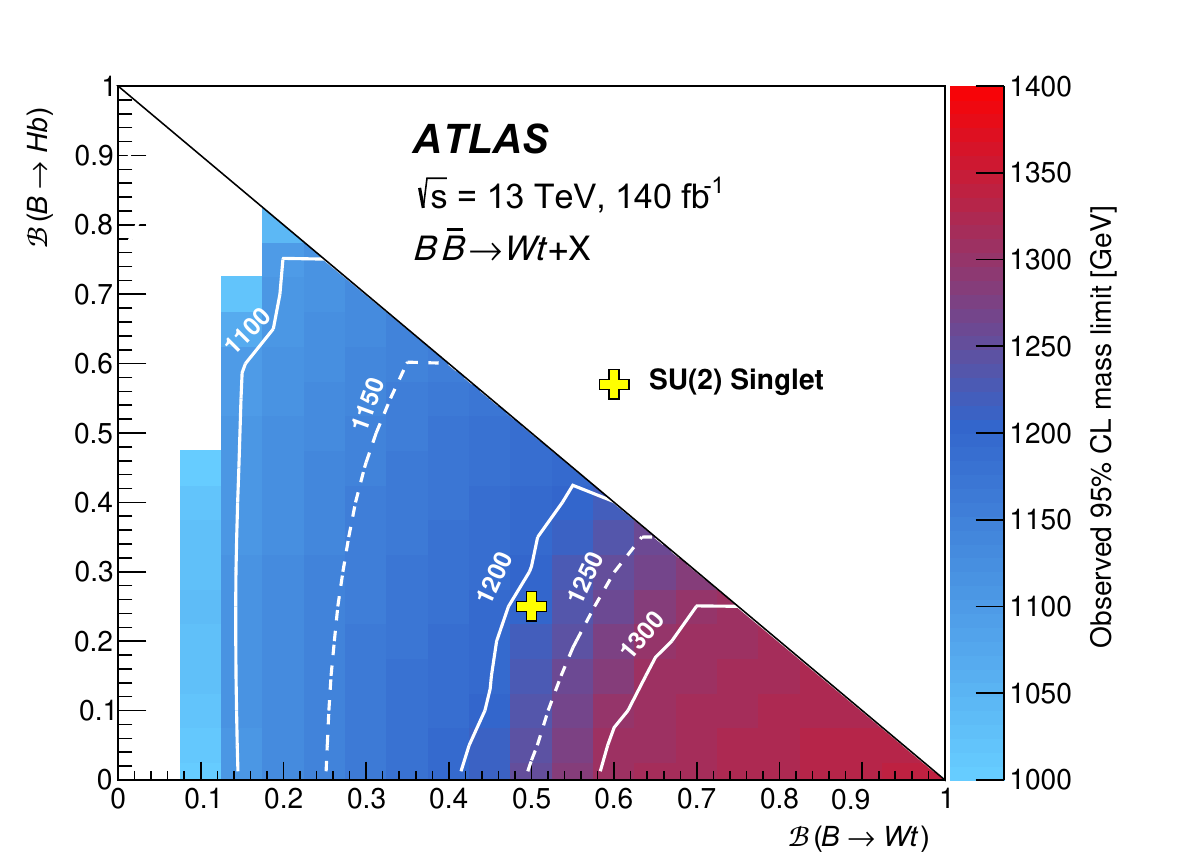}
\label{fig:BRScanB_b}
}
\caption{\protect\subref{fig:BRScanB_a} Expected and \protect\subref{fig:BRScanB_b} observed 95\% CL lower limits on the mass of the $B$-quark in the branching-ratio plane of \BR{B \to Wt} versus \BR{B \to Hb}. Contour lines are provided to guide the eye. The white region is due to the lower mass limit falling below \SI{1000}{\GeV}, the lowest signal mass considered in this search.  The yellow marker indicates the branching ratio for the SU(2) singlet $B$ scenario.}
\label{fig:BRScanB}
\end{figure}

In addition to the benchmark scenarios, other combinations of $T$ branching ratios are tested by reweighting the relative contributions of the three $T$-quark decay modes.
Figure~\ref{fig:BRScanT} shows the expected and observed lower limits on the $T$-quark mass as a function of $\BR{T \to Wb}$ and $\BR{T \to Ht}$.
For each point in the figure, the branching ratios for $T\to Zt$ decay are determined by the requirement $\BR{T \to Wb} + \BR{T \to Ht} + \BR{T \to Zt} = 1$.
Although the analysis is designed to search for \TTbar{}, it also has sensitivity to \BBbar{} production.
Figure~\ref{fig:BRScanB} shows the expected and observed lower limits on the $B$-quark mass as a function of $\BR{B \to Wt}$ and $\BR{B \to Hb}$, with $\BR{B \to Zb} = 1 - \BR{B \to Wt} - \BR{B \to Hb}$.
For the $\BR{B \to Wt}=1$ case, a lower limit on the $B$-quark mass of \MassLimitObsWbB  ~(\MassLimitExpWbB) are observed (expected).
Similarly to the $\BR{T \to Wb} = 1$ case, the $\BR{B \to Wt} = 1$ limits are applicable to the pair-production of vector-like $X$-quarks.

\FloatBarrier
 
\FloatBarrier
 
\section{Conclusion}
\label{sec:conclusion}

Pair-produced vector-like $T$-quarks are searched for in events with exactly one electron or muon, missing transverse momentum and jets using the full Run~2 ATLAS data set.
Final states compatible with the decay of a pair of heavy vector-like quarks are selected and a combined fit to the mass of the reconstructed vector-like quark in three control regions and two signal regions is made.
No significant excess over the background expectation is observed and 95\% CL upper limits are set on the vector-like quark cross-section and lower bounds are set on the vector-like quark mass.
The search is optimised for $\TTbar \to WbWb$, thus the most stringent limits are set for $\BR{T \to Wb} = 1$.
For this scenario, masses below \MassLimitObsWb{} (\MassLimitExpWb{}) are observed (expected) to be excluded at 95\% CL, which is the current strongest limit for this production and decay mode.
The limits for $\BR{T \to Wb} = 1$ also apply to a vector-like $Y$-quark with charge $-4/3$, which decays exclusively into $Wb$.
The observed (expected) lower mass limit for the weak isospin singlet $T$ model is \MassLimitObsSinglet{} (\MassLimitExpSinglet{}) and limits for other $T$-quark branching ratios are presented in the plane of $\BR{T\to Wb}$ vs. $\BR{T\to Ht}$.
The analysis was also used to set limits on the mass of $B$-quarks as a function of branching ratios.
This search improves the previous result that used 36\fb of ATLAS data by 350~\GeV for the scenario \BR{T \to Wb} = 1 and by 190~\GeV in the $T$ singlet case.


\section*{Acknowledgements}


We thank CERN for the very successful operation of the LHC and its injectors, as well as the support staff at
CERN and at our institutions worldwide without whom ATLAS could not be operated efficiently.
 
The crucial computing support from all WLCG partners is acknowledged gratefully, in particular from CERN, the ATLAS Tier-1 facilities at TRIUMF/SFU (Canada), NDGF (Denmark, Norway, Sweden), CC-IN2P3 (France), KIT/GridKA (Germany), INFN-CNAF (Italy), NL-T1 (Netherlands), PIC (Spain), RAL (UK) and BNL (USA), the Tier-2 facilities worldwide and large non-WLCG resource providers. Major contributors of computing resources are listed in Ref.~\cite{ATL-SOFT-PUB-2023-001}.
 
We gratefully acknowledge the support of ANPCyT, Argentina; YerPhI, Armenia; ARC, Australia; BMWFW and FWF, Austria; ANAS, Azerbaijan; CNPq and FAPESP, Brazil; NSERC, NRC and CFI, Canada; CERN; ANID, Chile; CAS, MOST and NSFC, China; Minciencias, Colombia; MEYS CR, Czech Republic; DNRF and DNSRC, Denmark; IN2P3-CNRS and CEA-DRF/IRFU, France; SRNSFG, Georgia; BMBF, HGF and MPG, Germany; GSRI, Greece; RGC and Hong Kong SAR, China; ISF and Benoziyo Center, Israel; INFN, Italy; MEXT and JSPS, Japan; CNRST, Morocco; NWO, Netherlands; RCN, Norway; MEiN, Poland; FCT, Portugal; MNE/IFA, Romania; MESTD, Serbia; MSSR, Slovakia; ARRS and MIZ\v{S}, Slovenia; DSI/NRF, South Africa; MICINN, Spain; SRC and Wallenberg Foundation, Sweden; SERI, SNSF and Cantons of Bern and Geneva, Switzerland; MOST, Taipei; TENMAK, T\"urkiye; STFC, United Kingdom; DOE and NSF, United States of America.
 
Individual groups and members have received support from BCKDF, CANARIE, CRC and DRAC, Canada; CERN-CZ, PRIMUS 21/SCI/017 and UNCE SCI/013, Czech Republic; COST, ERC, ERDF, Horizon 2020, ICSC-NextGenerationEU and Marie Sk{\l}odowska-Curie Actions, European Union; Investissements d'Avenir Labex, Investissements d'Avenir Idex and ANR, France; DFG and AvH Foundation, Germany; Herakleitos, Thales and Aristeia programmes co-financed by EU-ESF and the Greek NSRF, Greece; BSF-NSF and MINERVA, Israel; Norwegian Financial Mechanism 2014-2021, Norway; NCN and NAWA, Poland; La Caixa Banking Foundation, CERCA Programme Generalitat de Catalunya and PROMETEO and GenT Programmes Generalitat Valenciana, Spain; G\"{o}ran Gustafssons Stiftelse, Sweden; The Royal Society and Leverhulme Trust, United Kingdom.
 
In addition, individual members wish to acknowledge support from CERN: European Organization for Nuclear Research (CERN PJAS); Chile: Agencia Nacional de Investigaci\'on y Desarrollo (FONDECYT 1190886, FONDECYT 1210400, FONDECYT 1230812, FONDECYT 1230987); China: National Natural Science Foundation of China (NSFC - 12175119, NSFC 12275265, NSFC-12075060); Czech Republic: PRIMUS Research Programme (PRIMUS/21/SCI/017); European Union: European Research Council (ERC - 948254), Horizon 2020 Framework Programme (MUCCA - CHIST-ERA-19-XAI-00), European Union, Future Artificial Intelligence Research (FAIR-NextGenerationEU PE00000013), Italian Center for High Performance Computing, Big Data and Quantum Computing (ICSC, NextGenerationEU), Marie Sklodowska-Curie Actions (EU H2020 MSC IF GRANT NO 101033496); France: Agence Nationale de la Recherche (ANR-20-CE31-0013, ANR-21-CE31-0013, ANR-21-CE31-0022), Investissements d'Avenir Idex (ANR-11-LABX-0012), Investissements d'Avenir Labex (ANR-11-LABX-0012); Germany: Baden-Württemberg Stiftung (BW Stiftung-Postdoc Eliteprogramme), Deutsche Forschungsgemeinschaft (DFG - 469666862, DFG - CR 312/5-1); Italy: Istituto Nazionale di Fisica Nucleare (FELLINI G.A. n. 754496, ICSC, NextGenerationEU); Japan: Japan Society for the Promotion of Science (JSPS KAKENHI JP21H05085, JSPS KAKENHI JP22H01227, JSPS KAKENHI JP22H04944, JSPS KAKENHI JP22KK0227); Netherlands: Netherlands Organisation for Scientific Research (NWO Veni 2020 - VI.Veni.202.179); Norway: Research Council of Norway (RCN-314472); Poland: Polish National Agency for Academic Exchange (PPN/PPO/2020/1/00002/U/00001), Polish National Science Centre (NCN 2021/42/E/ST2/00350, NCN OPUS nr 2022/47/B/ST2/03059, NCN UMO-2019/34/E/ST2/00393, UMO-2020/37/B/ST2/01043, UMO-2021/40/C/ST2/00187); Slovenia: Slovenian Research Agency (ARIS grant J1-3010); Spain: BBVA Foundation (LEO22-1-603), Generalitat Valenciana (Artemisa, FEDER, IDIFEDER/2018/048), La Caixa Banking Foundation (LCF/BQ/PI20/11760025), Ministry of Science and Innovation (MCIN \& NextGenEU PCI2022-135018-2, MICIN \& FEDER PID2021-125273NB, RYC2019-028510-I, RYC2020-030254-I, RYC2021-031273-I, RYC2022-038164-I), PROMETEO and GenT Programmes Generalitat Valenciana (CIDEGENT/2019/023, CIDEGENT/2019/027); Sweden: Swedish Research Council (VR 2018-00482, VR 2022-03845, VR 2022-04683, VR grant 2021-03651), Knut and Alice Wallenberg Foundation (KAW 2017.0100, KAW 2018.0157, KAW 2018.0458, KAW 2019.0447); Switzerland: Swiss National Science Foundation (SNSF - PCEFP2\_194658); United Kingdom: Leverhulme Trust (Leverhulme Trust RPG-2020-004); United States of America: U.S. Department of Energy (ECA DE-AC02-76SF00515), Neubauer Family Foundation.


\clearpage

\printbibliography

\clearpage
 
\begin{flushleft}
\hypersetup{urlcolor=black}
{\Large The ATLAS Collaboration}

\bigskip

\AtlasOrcid[0000-0002-6665-4934]{G.~Aad}$^\textrm{\scriptsize 102}$,
\AtlasOrcid[0000-0002-5888-2734]{B.~Abbott}$^\textrm{\scriptsize 120}$,
\AtlasOrcid[0000-0002-1002-1652]{K.~Abeling}$^\textrm{\scriptsize 55}$,
\AtlasOrcid[0000-0001-5763-2760]{N.J.~Abicht}$^\textrm{\scriptsize 49}$,
\AtlasOrcid[0000-0002-8496-9294]{S.H.~Abidi}$^\textrm{\scriptsize 29}$,
\AtlasOrcid[0000-0002-9987-2292]{A.~Aboulhorma}$^\textrm{\scriptsize 35e}$,
\AtlasOrcid[0000-0001-5329-6640]{H.~Abramowicz}$^\textrm{\scriptsize 151}$,
\AtlasOrcid[0000-0002-1599-2896]{H.~Abreu}$^\textrm{\scriptsize 150}$,
\AtlasOrcid[0000-0003-0403-3697]{Y.~Abulaiti}$^\textrm{\scriptsize 117}$,
\AtlasOrcid[0000-0002-8588-9157]{B.S.~Acharya}$^\textrm{\scriptsize 69a,69b,m}$
% ...
\bigskip
\\

% ...
$^{29}$Physics Department, Brookhaven National Laboratory, Upton NY; United States of America.\\
% ...
$^{35}$$^{(a)}$Facult\'e des Sciences Ain Chock, R\'eseau Universitaire de Physique des Hautes Energies - Universit\'e Hassan II, Casablanca;$^{(b)}$Facult\'{e} des Sciences, Universit\'{e} Ibn-Tofail, K\'{e}nitra;$^{(c)}$Facult\'e des Sciences Semlalia, Universit\'e Cadi Ayyad, LPHEA-Marrakech;$^{(d)}$LPMR, Facult\'e des Sciences, Universit\'e Mohamed Premier, Oujda;$^{(e)}$Facult\'e des sciences, Universit\'e Mohammed V, Rabat;$^{(f)}$Institute of Applied Physics, Mohammed VI Polytechnic University, Ben Guerir; Morocco.\\
% ...
$^{49}$Fakult\"{a}t Physik , Technische Universit{\"a}t Dortmund, Dortmund; Germany.\\
% ...
$^{55}$II. Physikalisches Institut, Georg-August-Universit\"{a}t G\"ottingen, G\"ottingen; Germany.\\
% ...
$^{69}$$^{(a)}$INFN Gruppo Collegato di Udine, Sezione di Trieste, Udine;$^{(b)}$ICTP, Trieste;$^{(c)}$Dipartimento Politecnico di Ingegneria e Architettura, Universit\`a di Udine, Udine; Italy.\\
% ...
$^{102}$CPPM, Aix-Marseille Universit\'e, CNRS/IN2P3, Marseille; France.\\
% ...
$^{117}$Department of Physics, New York University, New York NY; United States of America.\\
% ...
$^{120}$Homer L. Dodge Department of Physics and Astronomy, University of Oklahoma, Norman OK; United States of America.\\
% ...
$^{150}$Department of Physics, Technion, Israel Institute of Technology, Haifa; Israel.\\
$^{151}$Raymond and Beverly Sackler School of Physics and Astronomy, Tel Aviv University, Tel Aviv; Israel.\\
% ...
$^{a}$ Also Affiliated with an institute covered by a cooperation agreement with CERN.\\
$^{b}$ Also at An-Najah National University, Nablus; Palestine.\\
$^{c}$ Also at Borough of Manhattan Community College, City University of New York, New York NY; United States of America.\\
$^{d}$ Also at Center for High Energy Physics, Peking University; China.\\
$^{e}$ Also at Center for Interdisciplinary Research and Innovation (CIRI-AUTH), Thessaloniki; Greece.\\
$^{f}$ Also at Centro Studi e Ricerche Enrico Fermi; Italy.\\
% ...
$^{m}$ Also at Department of Physics, King's College London, London; United Kingdom.\\
% ...

\end{flushleft}


\end{document}